\documentclass[a4paper,twocolumn,preprintnumbers,citeautoscript,aps,prl,longbibliography,superscriptaddress,floatfix,nofootinbib]{revtex4-2}
\usepackage{amsmath,amsfonts,amssymb,graphics}
\usepackage[normalem]{ulem}
\usepackage{url}
\usepackage{graphicx}

\usepackage{mathrsfs}
\usepackage{bbm}
\usepackage{pifont}
\usepackage{braket}
\usepackage{siunitx}
\usepackage{slashed}

\usepackage{mathtools}
\usepackage{xspace}
\usepackage{booktabs}
\usepackage{csquotes}
\usepackage[dvipsnames,svgnames]{xcolor}
\usepackage{tikz}
\usetikzlibrary{positioning,shapes,arrows}
\usetikzlibrary{decorations.pathmorphing}
\usetikzlibrary{decorations.markings}
\usepackage{standalone}

\usepackage{url}
\usepackage[bookmarks=false, breaklinks]{hyperref} 

\definecolor{nicered}{rgb}{0.5,.0,.0}
\definecolor{darkblue}{rgb}{0,.1,.9}
\definecolor{lightblue}{rgb}{0,.1,.6}
\definecolor{applegreen}{rgb}{0.55, 0.71, 0.0}
\definecolor{darkgreen}{rgb}{0.0, 0.2, 0.13}
 
\hypersetup{colorlinks, citecolor=darkgreen ,linkcolor=nicered, urlcolor= lightblue}

\begin{document}
\title{Minimal solution to the axion isocurvature problem from a nonminimal coupling}
\author{Maximilian Berbig}
\email{berbig@ific.uv.es}
\affiliation{Departament de Física Teòrica, Universitat de València, 46100 Burjassot, Spain}
\affiliation{Instituto de Física Corpuscular (CSIC-Universitat de València), Parc Científic UV, C/Catedrático José Beltrán, 2, E-46980 Paterna, Spain}

\date{\today}

\begin{abstract}
The main limitation for preinflationary breaking of Peccei-Quinn (PQ)  symmetry is the upper bound on the Hubble rate during inflation from axion isocurvature fluctuations. This leads to a tension between high scale inflation and QCD axions with  grand unified theory (GUT) scale decay constants, which reduces the potential  for a detection of tensor modes  at next generation cosmic microwave background (CMB) experiments. We propose a mechanism that explicitly breaks PQ symmetry via nonminimal coupling to gravity, that lifts the axion mass above the Hubble scale during inflation and has negligible impact on today's axion potential. The initially heavy axion gets trapped at an intermediate minimum during inflation given by the phase of the nonminimal coupling, before it moves to its true $CP$-conserving minimum after inflation.
During this stage it undergoes coherent oscillations around an adiabatically decreasing minimum, which slightly dilutes the axion energy density, while still being able to explain the observed dark matter relic abundance. This scenario can  be tested by the combination of next generation CMB surveys like CMB-S4 and LiteBIRD   with haloscopes such as ABRACADABRA, DMRadio or CASPEr-Electric.
\end{abstract}

\maketitle

\textbf{Introduction.\textemdash}The Peccei-Quinn (PQ) mechanism \cite{PhysRevLett.38.1440,Peccei:1977ur} relying on an anomalous, global $\text{U}(1)_\text{PQ}$ symmetry is considered to be one of the most attractive solutions to the strong $CP$-problem and the accompanying (super-)light pseudoscalar known as the axion \cite{PhysRevLett.40.223,PhysRevLett.40.279} is a promising dark matter (DM) candidate, see Refs.~\cite{DiLuzio:2020wdo,OHare:2024nmr} for  reviews. Depending on whether PQ symmetry is broken before \cite{Abbott:1982af,Dine:1982ah,Preskill:1982cy} or after the visible part of inflation (see Refs.~\cite{Redi:2022llj,Gorghetto:2023vqu} for the case in between), different axion cosmologies emerge. 

The spontaneous breaking of $\text{U}(1)_\text{PQ}$ with a vacuum expectation value proportional to $f_a$ occurs before inflation and it is never restored by either quantum or thermal fluctuations as long as 
\begin{align}\label{eq:max}
    f_a > \text{Max}\left[\frac{H_I}{2\pi}, T_\text{max}, T_\text{RH}\right].
\end{align}
Here the first term denotes the Gibbons-Hawking temperature \cite{PhysRevD152738} during inflation with a Hubble rate $H_I$, $T_\text{max}$ the maximum temperature during inflationary reheating \cite{Giudice:2000ex,Kolb:2003ke} and $T_\text{RH}$ the temperature of the radiation bath after reheating.\footnote{$T_\text{max}$ is typically expected to be larger than $T_\text{RH}$, but exceptions to this common lore exist \cite{Co:2020xaf}.} 
In this preinflationary scenario there is no threat from topological defects and the massless axion $a  \equiv \theta f_a$ develops long wavelength quantum fluctuations  $\delta \theta \simeq  H_I /(2\pi f_a ) $ during the quasi-de Sitter phase, that give rise to a scale invariant power spectrum with an amplitude proportional to $H_I/(2\pi^2)$. Cosmic microwave background (CMB) data, which constrain the fraction of the amplitude of isocurvature to adiabatic fluctuations to be $\alpha<0.038$ \cite{Planck:2018jri}, demand that the misalignment angle after inflation $\theta_I$ is larger than its fluctuation $\theta_I^2 \gg \delta \theta^2$ \cite{Hamann:2009yf}.
Assuming that the misalignment mechanism \cite{Preskill:1982cy,Abbott:1982af,Dine:1982ah} makes up all of the dark matter (DM) relic density, the limit on $\alpha$ implies the bound 
\begin{align}
    H_I < \SI{2.98e11}{\giga\electronvolt}\cdot \theta_I \left(\frac{f_a}{10^{16}\text{GeV}}\right),
\end{align}
where $\theta_I$ has to be fixed for a given  $f_a$ to reproduce the DM relic abundance (e.g. $\theta_I\ll1$ for $f_a\simeq 10^{16}\text{GeV}$).
For $\theta_I \simeq \pi$, required to fit the relic abundance in the regime $f_a\ll 10^{12}\;\text{GeV}$ with the help of anharmonicity effects \cite{Turner:1985si,Lyth:1991ub}, the limit  on $H_I$ gets stronger as it depends on $\theta_I-\pi$ instead \cite{Visinelli:2009zm,Kobayashi:2013nva,GrillidiCortona:2015jxo}.
PQ breaking at the grand unified theory (GUT) scale of $10^{16}\text{GeV}$, which is expected in GUTs \cite{Wise:1981ry,Lazarides:1981kz} and some string theory constructions \cite{Choi:1985je,Svrcek:2006yi,Arvanitaki:2009fg},  will be probed by the haloscopes ABRACADABRA (phase 1) \cite{Kahn:2016aff}, DMRadio \cite{DMRadio:2022pkf,DMRadio:2022jfv} and CASPEr-Electric (phase 2) \cite{Graham:2013gfa,Budker:2013hfa,JacksonKimball:2017elr}.
This regime for $f_a$ implies a tensor to scalar ratio $r$ for single-field, slow-roll inflation scaling as  $r\simeq 8.75\times 10^{-7} (H_I/(\theta_I \cdot \SI{3e11}{\giga\electronvolt}))^2$, and  would thus be incompatible with observable gravitational waves from inflation at next generation experiments  aiming for $r\simeq 10^{-3}$, such as CMB-S4 \cite{CMB-S4:2016ple} or  LiteBIRD \cite{Matsumura:2013aja};
the current bound from BICEP/Keck reads $r<0.036$~\cite{BICEP:2021xfz}.
The tension between preinflationary PQ breaking and large $H_I$ is known as the \enquote{axion isocurvature problem}.

We propose a mechanism inducing a nonzero axion mass during inflation similar to previous proposals involving  either a modified running of QCD \cite{Dvali:1995ce,Jeong:2013xta,Koutsangelas:2022lte}, PQ breaking from new confining groups, gravity \cite{Takahashi:2015waa} or  hidden monopoles \cite{Nomura:2015xil,Kawasaki:2017xwt} via the Witten effect \cite{Witten:1979ey}. 
Our idea involves no additional degrees of freedom as it relies on a gravitational effect similar to Refs.~\cite{Folkerts:2013tua,Takahashi:2015waa}.
Other avenues to solve the axion isocurvature problem involve a field excursion of the PQ breaking field's radial mode much larger than  $f_a$ \cite{Linde:1991km,Choi:2014uaa,Chun:2014xva,Higaki:2014ooa,Fairbairn:2014zta,Nakayama:2015pba,Harigaya:2015hha,Kearney:2016vqw},  restoring PQ symmetry via a coupling to the inflaton \cite{Bao:2022hsg} or  producing isocurvature at late times during inflation, so it appears only on  scales out of reach of the CMB and the Lyman-$\alpha$ forest \cite{Kawasaki:2017ycl,Rosa:2021gbe,Redi:2022llj}.

We begin by presenting our mechanism and then elaborate on the implications for the axion quality problem. Further we discuss the impact on the evolution of the axion after inflation, before summarizing.

\begin{figure*}[t]
    \centering
    \includegraphics[width=0.45\textwidth]{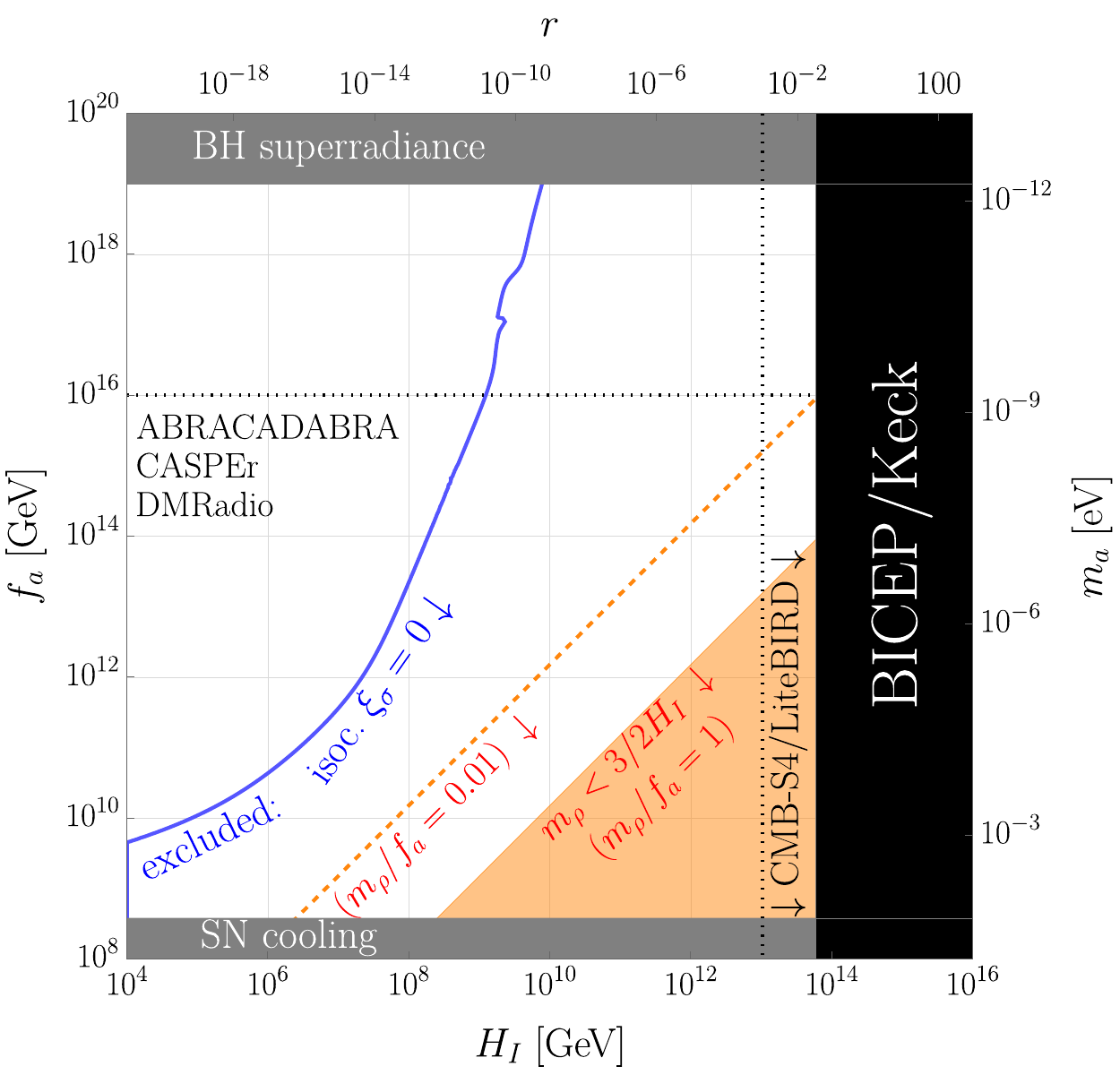}\
    \hspace{0.03\textwidth}
    \includegraphics[width=0.45\textwidth]{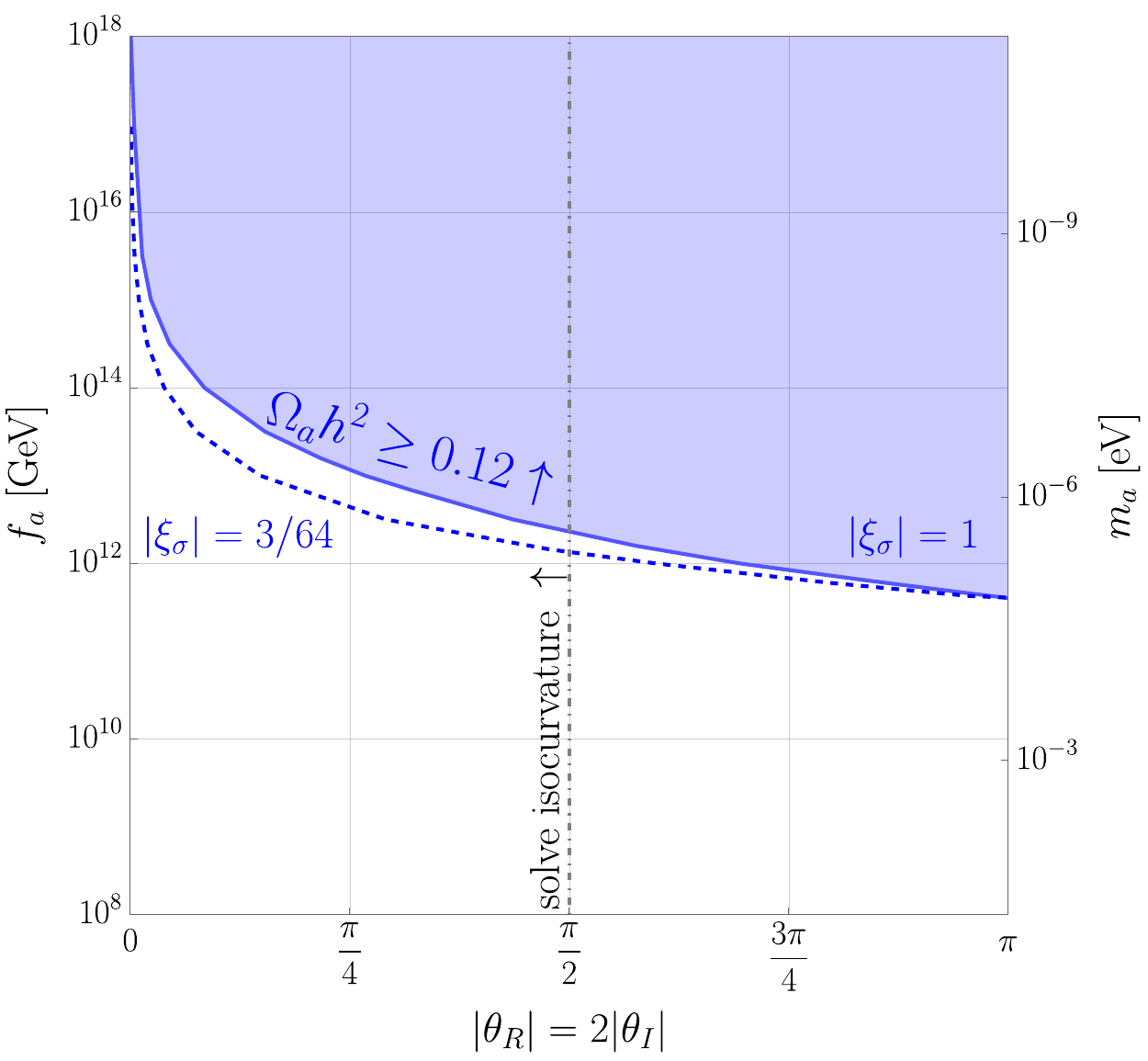}
    \caption{Left: Parameter space for our mechanism to solve the axion isocurvature problem. If the non minimal coupling $\xi_\sigma$ vanishes, the entire region to the right of the blue line would be excluded. The kink at $f_a\simeq 10^{17}\text{GeV}$ arises due to our analytical approximation for the temperature dependence of $m_a$ in Eq.~\eqref{eq:maT}. For $\xi_\sigma\neq 0$ the problem is solved in the entire  white area outside of  the orange region, which  depicts the bound in Eq.~\eqref{eq:bound1}.
    We do not depict the region for $f_a<H_I/(2\pi)$, as it is contained in the lower orange area for $m_\rho=f_a$.  The constraint on black hole (BH) superradiance stems from Ref.~\cite{Cardoso:2018tly} and the supernova (SN) cooling limit is from Refs.~\cite{Fischer:2016cyd,Carenza:2019pxu}.
    Right: Impact of the adiabatic suppression for $\xi_\sigma\neq 0$ on the parameter space  reproducing the dark matter relic abundance via coherent oscillations of the QCD axion for two representative values of $|\xi_\sigma|=1,\; 3/64$. The isovurvature problem can be solved for $|\theta_R|<\pi/2$ (within the bounds of Eq.~\eqref{eq:xi}) and larger values of $|\theta_I|$ not determined by $|\theta_R|$ could be possible, if there was an epoch of kination after inflation (see the discussion before  Eq.~\eqref{eq:VQCD}). The curve for $\xi_\sigma=0$ was not displayed as it mostly overlaps with the one for $|\xi_\sigma|=3/64$.}
    \label{fig:isoc-1} 
\end{figure*}

\textbf{The mechanism.\textemdash}We introduce an explicit PQ symmetry breaking nonminimal coupling of the singlet field $\sigma \equiv (\rho+f_a) \exp{(i \theta)}/\sqrt{2}$, with $\rho$ being the radial mode and  $\theta \equiv a/ \rho$ denoting the axion, to gravity
\begin{align}\label{eq:op}
    V(\sigma)\supset  \xi_\sigma R \sigma^2 + \text{h.c.}=  \left|\xi_\sigma\right| R  \rho^2 \cos{\left(2\theta - \theta_R\right)},
\end{align}
where $\theta_R = - \text{Arg}\left[\xi_\sigma\right]\in [-\pi,\pi]$ denotes a physical phase, since a PQ transformation would just move it into the QCD instanton induced term in Eq. \eqref{eq:VQCD} of one of the next sections.
A similar operator corresponding to $\rho$ being fixed at the Planck scale was proposed in the Appendix of Ref. \cite{Takahashi:2015waa} and operators of the form $R \sigma^2$ appear in Refs. \cite{Hashimoto:2021xgu,Barenboim:2024akt,Barenboim:2024xxa}. 
Couplings of massive vectors to curvature were studied in Refs.~\cite{Himmetoglu:2008zp,Himmetoglu:2009qi,Karciauskas:2010as,Arias:2012az,Alonso-Alvarez:2019ixv,Nakayama:2019rhg,Hell:2024xbv}.
There is no exact symmetry to forbid the above operator, and, in general, one expects, that scalar couplings to curvature are induced radiatively for self-interacting scalars in curved spacetimes,  see e.g. Refs.~\cite{Birrell:1982ix,Parker:2009uva}. For simplicity we assume that the coupling of  $\xi  R |\sigma|^2$ satisfies $|\xi| \ll |\xi_\sigma|$ and that inflation is sourced by a different scalar field $\varphi$.  If this field has a nonminimal coupling $\xi_\varphi$, we require $\left|\xi_\sigma\right| < \xi_\varphi$.
Inflation from the radial mode of the PQ scalar was covered in Refs.~\cite{Fairbairn:2014zta,Ballesteros:2016xej,Kearney:2016vqw,Boucenna:2017fna,Hamaguchi:2021mmt,DalCin:2023uai} and hybrid inflation from the radial and angular modes in Refs.~\cite{McDonough:2020gmn,Barenboim:2024akt,Barenboim:2024xxa}.
We can neglect the effect of the transformation from the Jordan frame to the Einstein frame as long as $ \left|\xi_\sigma\right| \cos{\left(2\theta-\theta_R\right)} f_a^2 \ll M_\text{Pl.}^2$. Since the cut-off scale of quantum gravity is approximately given by $M_\text{Pl.}/\sqrt{\left|\xi_\sigma\right|}$ \cite{Lerner:2009na,Hertzberg:2010dc,Burgess:2010zq} we conservatively require $\left|\xi_\sigma\right|<1$.
The Ricci scalar $R$ is found in terms of the Equation of state parameter $\omega$ of the dominant component driving the expansion of the Universe as 
\begin{align}
    R= - 3 ( 1- 3 \omega) H^2,
\end{align}
and from this we obtain the effective axion mass during inflation ($\omega \simeq - 1$ and a constant $H_I$  during the $N_e\simeq 50-60$ e-folds of visible inflation) by expanding in $a\ll \braket{\rho}=f_a$ 
\begin{align}\label{eq:central}
    M_a^2 = 48 \left|\xi_\sigma\right|  \cos{\left(\theta_R\right)} H_I^2.
\end{align}
Note that unlike the construction in Ref.~\cite{Takahashi:2015waa}  this mass is independent of the axion decay constant $f_a$ due to the quadratic coupling to the radial mode in Eq.~\eqref{eq:op}.
$H_I$-dependent masses during inflation can also arise from direct couplings to the inflaton, since $H_I^2 \sim V(\varphi)/M_\text{Pl.}^2$. However our approach automatically avoids potentially dangerous  modifications of the inflaton potential $V(\varphi)$ from the coupling of $\varphi$ to $\sigma$, which could spoil its flatness required for successful inflation \cite{Kearney:2016vqw}. The axion  develops  only suppressed quantum fluctuations on superhorizon scales, if it is too heavy  to be excited from the inflationary vacuum \cite{Chen:2009zp}
\begin{align}\label{eq:noiso}
    M_a > \frac{3}{2} H_I.
\end{align} 
The  isocurvature power spectrum has an amplitude reduced by a factor of  $H_I/M_a$ \cite{Chung:2004nh,Garcia:2023qab} and becomes  blue tilted (suppressed for large wavelengths) with a spectral index of 3 \cite{Liddle:1999pr}, so that the CMB limits are evaded. The amplitude for the production of axions is suppressed by $e^{-\frac{\pi M_a}{H_I}}$ \cite{Li:2019ves,Sou:2021juh}.
A tachyonic mass squared for the axion would lead to perturbations that grow exponentially in time \cite{Linde:2001ae}, which is why we demand $M_a^2>0$, corresponding to $|\theta_R| \in [0, \pi/2)$.
Our description in terms of a pseudo-Nambu-Goldstone boson holds for $M_a^2 < m_\rho^2 \equiv 2 \lambda_\sigma f_a^2$ \cite{Co:2022aav}, where the quartic potential $\lambda_\sigma (|\sigma|^2 -f_a^2/2)^2$ was assumed. This automatically ensures that the minimum of the radial mode $\braket{\rho} =f_a$ is not shifted by the presence of the $R$-dependent term, and thus, there is no field excursion in the radial direction.
Both conditions together enforce the range 
\begin{align}\label{eq:xi}
    \frac{3}{64} <  \left|\xi_\sigma\right| \cos{\left(\theta_R\right)} < \frac{1}{48} \frac{m_\rho^2}{H_I^2}.
\end{align}
Demanding that the upper limit of the above is larger than the lower limit implies that $ m_\rho >  3/2 H_I$, which is the condition for the absence isocurvature fluctuations in the radial direction, \footnote{Hence neither the radial mode nor the axion can act as a curvaton \cite{Linde:1996gt,Enqvist:2001zp,Lyth:2001nq,Moroi:2001ct}.} or equivalently
\begin{align}\label{eq:bound1}
 f_a > \SI{9e13}{\giga\electronvolt} \cdot \left(\frac{f_a}{m_\rho}\right) \left(\frac{H_I}{\SI{6e13}{\giga\electronvolt}}\right).
\end{align}
Since $\rho$ is heavier than both the axion and  $H_I$, it will sit at its minimum $f_a$ during inflation. Consequently there can be no restoration of the PQ symmetry via parametric resonance from radial oscillations \cite{Tkachev:1995md,Kofman:1995fi,Kasuya:1996ns,Kasuya:1997ha,Kasuya:1998td,Tkachev:1998dc,Kasuya:1999hy} and we do not have to worry about the contribution to the axion mass from the relaxation period of $\rho$ \cite{Ballesteros:2021bee}. 
Our setup assumes that the PQ fields are spectators during inflation and their potential in Eq.~\eqref{eq:op}  does not drive inflation for 
\begin{align}
    \frac{\theta f_a}{M_\text{Pl.}} <\sqrt{\frac{1}{32 \pi\left|\xi_\sigma\right|  \cos{\left(\theta_R\right)}}}<0.46,
\end{align}
where we  used the lower bound from \eqref{eq:xi} in the last inequality. Applying the same condition to the quartic  potential of the radial mode   leads to  no constraint as the potential  $\lambda_\sigma (\rho^2 -f_a^2)^2 /4$ vanishes once the heavy radial mode relaxes to its minimum $\braket{\rho}=f_a$. 
We depict the parameter space for our mechanism singled out by the previous constraints in the left panel of fig.~\ref{fig:isoc-1}. The conventional isocurvature limits for $\xi_\sigma= 0$ were computed from the analytical expression for the relic abundance in Eq.~\eqref{eq:relic}  following Ref.~\cite{Visinelli:2009zm}.

\textbf{Quality problem.\textemdash}Here, we do not attempt to answer, why the operator in Eq.~\eqref{eq:op} is realized instead of any term like $\sigma^d M_\text{pl.}^{4-d}$ usually expected from nonperturbative quantum gravitational effects discussed in the context of the axion quality problem \cite{Georgi:1981pu,Dine:1986bg,Coleman:1989zu,Abbott:1989jw,Holman:1992us,Kamionkowski:1992mf,Barr:1992qq,Ghigna:1992iv,Alonso:2017avz}. Reference~\cite{Kallosh:1995hi} showed that such an explicit violation of PQ symmetry can be suppressed by either a large Euclidean wormhole action, a large topological wormhole action from the Gauss-Bonnet term, or both. Of course this begs the question, if the operator in Eq. \eqref{eq:op}
would still be allowed, and why it is realized instead of e.g. $M_\text{Pl.}^2 \sigma^2$  (or $R |\sigma|^2$) to begin with.\footnote{Due to the unbroken residual $\mathcal{Z}_2$ of Eq.~\eqref{eq:op} there is no symmetry that prevents the dangerous operator $M_\text{Pl.}^2 \sigma^2$ that spoils the PQ mechanism. Such a term might be reintroduced via quantum corrections involving graviton loops from the coupling in \eqref{eq:op}. Here we have to assume that the resulting coupling is negligibly small, which constitutes a serious caveat.}
 
In the absence of a definitive solution to the quality problem we assume the absence of all $M_\text{Pl.}$-dependent operators and focus on the phenomenological impact of Eq.~\eqref{eq:op}.
This coupling poses no threat in terms of the quality problem, as the modification of the axion potential during the current era of accelerated expansion ($\omega\simeq -1$) depends on the tiny Hubble rate today $H_0$: The minimum $\theta_0\simeq 0$ \footnote{The Vafa-Witten theorem \cite{Vafa:1984xg} ensures that the minimum for a vector-like gauge theory would be at $\theta_0 =0$. Since the electroweak Standard Model involves chiral quarks, one expects that the actual minimum  lies at $\theta_0\simeq G_F^2 f_\pi^4 J\simeq 10^{-18}$ \cite{Georgi:1986kr,DiLuzio:2020wdo} in terms of Fermi's constant $G_F$, the pion decay constant $f_\pi$ and the Jarlskog invariant $J$ encoding the $CP$-violation in the quark sector.} of the combined potential with a contribution from QCD instantons, see Eq.~\eqref{eq:VQCD} in the following section, with a zero-temperature axion mass of \cite{Gorghetto:2018ocs}
\begin{align}
    m_a &\equiv \SI{5.7}{\micro\electronvolt} \left(\frac{10^{12}\; \text{GeV}}{f_a}\right),\label{eq:ma}
\end{align}
is  shifted to $|\theta_0| < 10^{-10}$ as long as 
\begin{align}\label{eq:qualbound}
 f_a  <  \frac{\SI{7e33}{\giga\electronvolt}}{\sqrt{\left|\xi_\sigma\right|   \sin{\left(\theta_R\right)}}} \left(\frac{70 \frac{\text{km}}{\text{s}} \text{Mpc}^{-1}}{H_0} \right) \sqrt{\frac{|\theta_0|}{10^{-10}} },
\end{align}
where we used that $m_a\gg M_a(H_0)$. 
Since Earth is in a gravitationally bound state the local value of curvature might actually be larger than $12 H_0^2$. A crude estimate can be obtained with  $R=8\pi \rho_e/M_\text{Pl.}^2 = \mathcal{O}(10^{-36}\;\text{eV}^2)$ in terms  of the energy densities of Earth or the Sun \textcolor{red}{$\rho_e$}, which are of comparable magnitude. Using this estimate the bound in Eq.~\eqref{eq:qualbound} is tightened to $\SI{2e19}{\giga\electronvolt}$.

\begin{figure*}[t]
    \centering
    \includegraphics[width=0.5\textwidth]{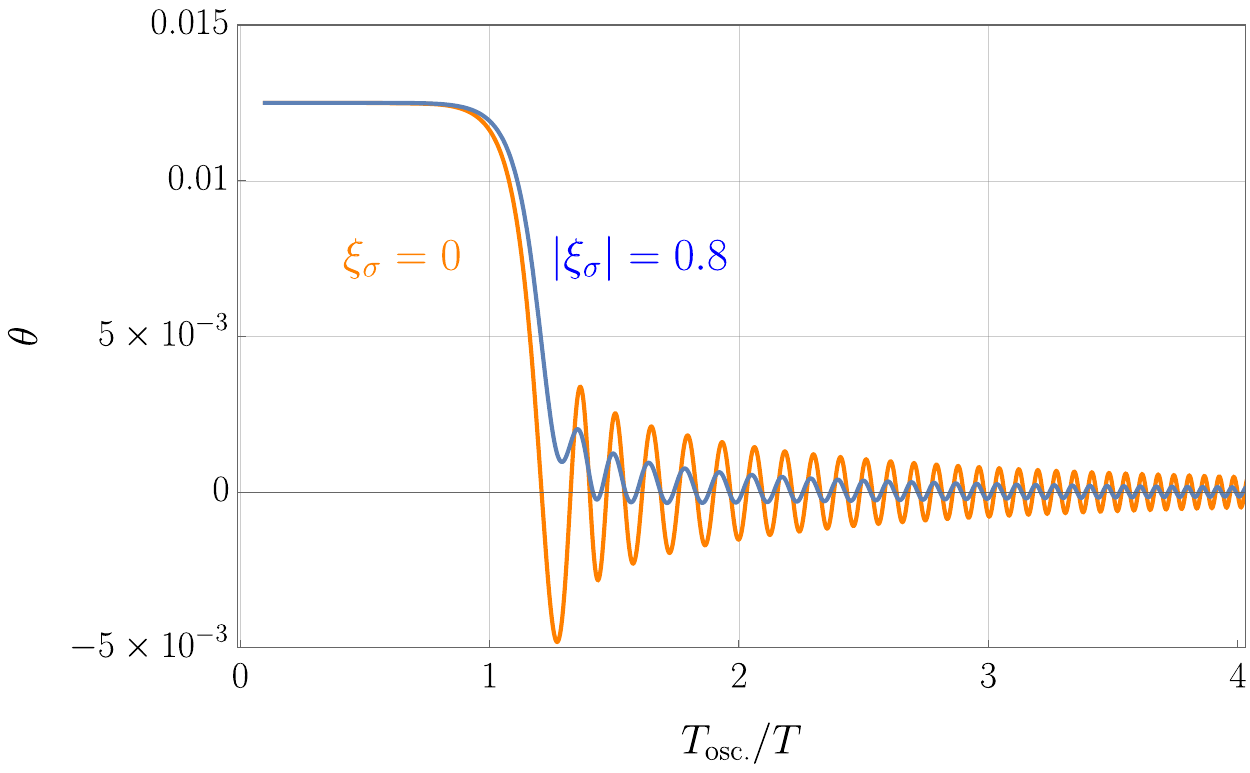}
    \hspace{0.03\textwidth}
    \includegraphics[width=0.45\textwidth]{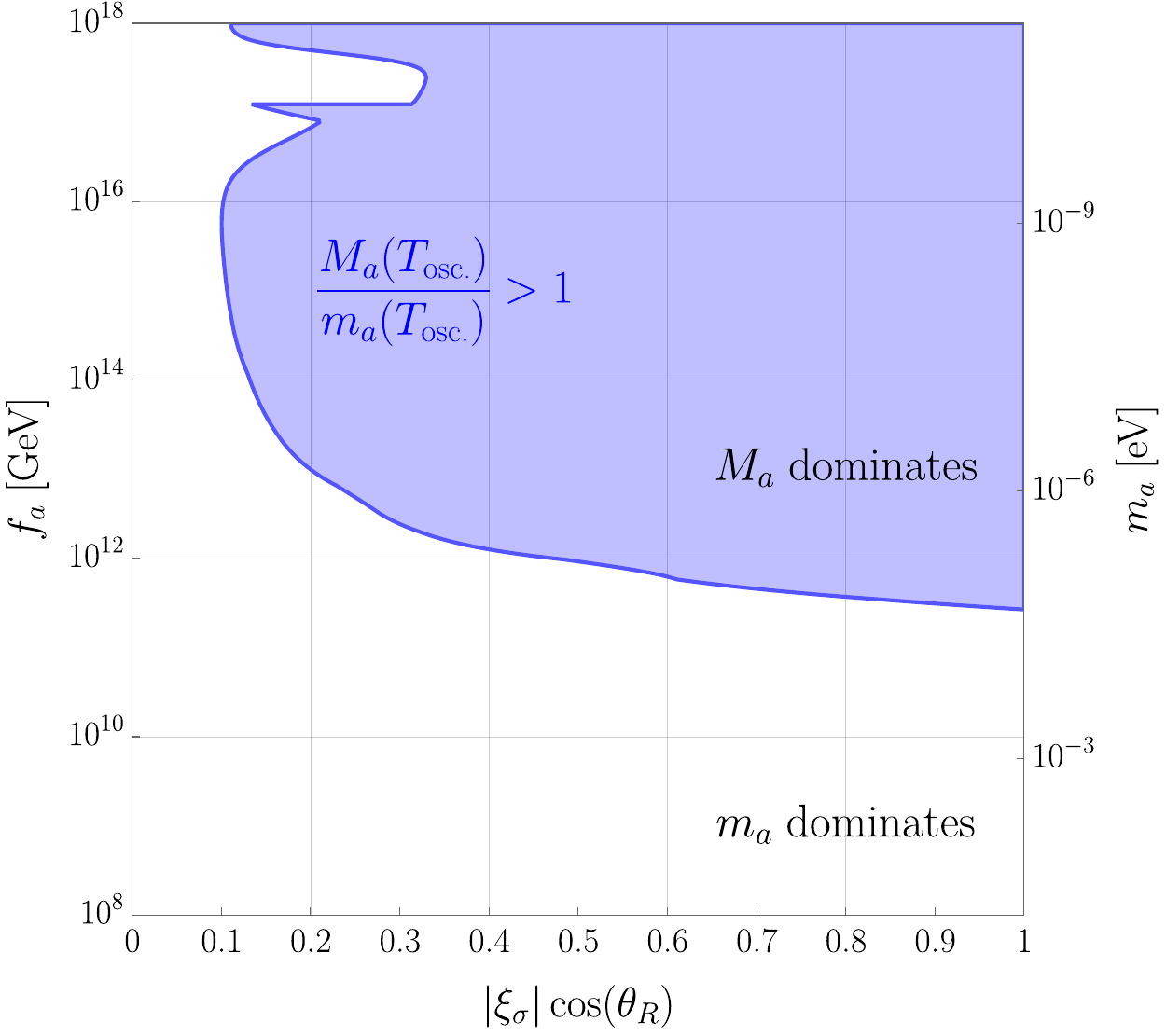}
    \caption{Left:  Evolution of the axion misalignment angle $\theta$ with $f_a = 10^{16}\;\text{GeV}$ and $\theta_R=2\theta_I=0.025$ for $\xi_\sigma = 0$ (orange) and $|\xi_\sigma|=0.8$ (blue). The oscillations for $\xi_\sigma = 0$ start at $T_\text{osc.}$, defined via $m_a(T_\text{osc.})=3 H(T_\text{osc.})$, see Eqs.~\eqref{eq:ma}, \eqref{eq:maT}. Taking $\xi_\sigma \neq 0$ both slightly delays the onset of the oscillation and reduces its amplitude. Right: Parameter space  in which the  Hubble dependent mass $M_a(T)= c_a(T) H(T)$ (see Eq.~\eqref{eq:adiab}) is larger than the QCD axion mass $m_a(T)$  evaluated at $T_\text{osc.}$, when the oscillations would start for $M_a=0\;(\xi_\sigma = 0)$. The blue (white) region indicates that the onset of oscillations is determined by $M_a(T)$ ($m_a(T)$) and very roughly corresponds to  $f_a> \SI{3e11}{\giga\electronvolt} \;(f_a< \SI{3e11}{\giga\electronvolt})$. Note that the kinks around $f_a\simeq 10^{17}\text{GeV}$ arise due to our analytical approximation for the temperature dependence of $m_a$ in Eq.~\eqref{eq:maT}.}
    \label{fig:osc-1}
\end{figure*}

\textbf{Post-inflationary evolution.\textemdash}The unknown initial position of the axion field will be denoted by $\theta_i \in [-\pi,\pi]$. In the standard case of preinflationary PQ breaking, one find that the position of the axion after inflation is $\theta_I=\theta_i$, and it follows that the relic abundance from the misalignment mechanism scaling as $\theta_I^2$ depends on initial conditions. In our case the axion will relax to the minimum of its Hubble-dependent potential over a timescale $\tau_\text{relax}\simeq 1/M_a$ which is shorter than the age of the Universe at the end of inflation $\tau \simeq 1/H_I$  due to the requirement for the absence of axion isocurvature in Eq.~\eqref{eq:noiso}. After inflation we find $\theta_I = \theta_R/2$ and the initial condition for the axion is replaced by a Lagrangian parameter, that is, at least in principle, calculable from a given UV completion inducing the operator in Eq.~\eqref{eq:op}. 
This early trapping of the axion in an intermediate vacuum is reminiscent of the scenarios in Refs.~\cite{Nakagawa:2020zjr,DiLuzio:2021gos,Jeong:2022kdr}. If the coupling to gravity conserves $CP$($\theta_R=0$) one finds after $N_e$ e-folds of inflation  that $|\theta_I| =|\theta_i| e^{-3N_e/2} \lesssim 3\times10^{-13}$ \cite{Buen-Abad:2019uoc}, where we used $|\theta_i|\leq \pi$ and $N_e= 60$. However as $\theta_R =0$ effectively implies $|\theta_I|\simeq0$ and thus no coherent oscillations after inflation, we focus on $\theta_R \neq 0$.

Cosmic strings from the spontaneous breaking of PQ symmetry \cite{Kibble_1976,PhysRevD.26.435,KIBBLE1980183}  with $f_a > H_I$ are diluted by the exponential expansion during inflation. The coupling in Eq.~\eqref{eq:op} breaks $\text{U}(1)_\text{PQ}$ down to $\mathcal{Z}_2$. One finds degenerate minima for the given potential at $\theta_I =\theta_R/2 + n\pi, \; n\in \mathbb{Z}$, which leads to  domain walls (DW), that are also inflated away.
Due to the absence of quantum fluctuations in the axion direction we do not expect exponentially large DWs, as they require $|\delta \theta| \gg \pi$ \cite{Linde:1990yj}.

During reheating (assuming $\omega=0$) as long as  the Hubble-induced mass remains nonzero, we do not expect thermal fluctuations to displace the axion (the cases of an oscillating Ricci scalar or a nonadiabatic transition to reheating  are discussed in the next section above Eq.~\eqref{eq:fluct}.). Even if the axion potential was absent, the average displacement would be at most $\Delta \theta \simeq T_\text{RH}/f_a$ 
\cite{Koutsangelas:2022lte}, which is small under the assumption in Eq.~\eqref{eq:max}. 
We assume that the radial mode undergoes fast decays to particles other than the axion, to avoid intermediate matter domination \cite{Nelson:2018via} and the overproduction of axion dark radiation \cite{Co:2019jts}.

If there is an epoch with a stiff equation of state $1/3\leq\omega \leq 1$,
also known as kination \cite{Spokoiny:1993kt,Joyce:1996cp,Ferreira:1997hj,Co:2021lkc,Gouttenoire:2021wzu,Gouttenoire:2021jhk}, the sign of $\omega$ changes  compared to inflation. Consequently one finds  $M_a^2<0$  during this era \cite{Bettoni:2018utf}, implying that the minimum at $|\theta_R|$ becomes a maximum and that the axion rolls down to one of the two neighboring minima. 
After kination, when the sign of $M_a^2$ flips again, this minimum also turns into a maximum and  the axion will start rolling down from there.  
This sequence of events changes the dependence of $\theta_I$ on $\theta_R$,  
and a dedicated analysis is required to treat this case. In general  starting at   $|\theta_I|=|\theta_R|/2=\pi$ is not possible for our quadratic potential, as this would require $|\theta_R|=2\pi$.

We briefly summarize the conventional misalignment mechanism \cite{Preskill:1982cy,Abbott:1982af,Dine:1982ah} during radiation domination: 
The QCD induced potential reads 
\begin{align}\label{eq:VQCD}
        V_\text{QCD} = m_a(T)^2 f_a^2 \left( 1- \cos(\theta)\right)
\end{align}
and the temperature dependence of the axion mass from the QCD susceptibility was computed in \cite{Borsanyi:2016ksw} using lattice methods. We use the analytical fitting formulas from Refs. \cite{Hertzberg:2008wr,Arias:2021rer,Arias:2022qjt} 
\begin{align}\label{eq:maT}
    m_a(T) = m_a
    \begin{cases}
        1 \quad &\text{for}\quad  T\leq\SI{150}{\mega\electronvolt}\\
         0.02 \left(\frac{\Lambda_\text{QCD}}{T}\right)^4 \quad &\text{for}\quad  T > \SI{150}{\mega\electronvolt}
    \end{cases},
\end{align}
with $\Lambda_\text{QCD} \simeq \SI{400}{\mega\electronvolt}$ and $m_a$ defined in Eq.~\eqref{eq:ma} of the main text. 
The axion field remains frozen in its postinflationary position of $\theta_I$, before it starts to oscillate at $T_\text{osc.}$, defined via  $m_a(T_\text{osc.})\simeq 3 H(T_\text{osc.})$, and relaxes to its $CP$-conserving minimum $\theta\simeq 0$. Oscillations of the axion at $T_\text{osc.}\leq 0.02^\frac{1}{4} \Lambda_\text{QCD}\simeq \SI{150}{\mega\electronvolt}$ correspond to $f_a \geq \SI{1.24e17}{\giga\electronvolt}$.
 The vacuum realignment  \cite{Preskill:1982cy,Abbott:1982af,Dine:1982ah} of the axion leads to a dark matter relic density of
\begin{align}\label{eq:relic}
   \Omega_a h^2 \simeq 0.12 
    \begin{cases}
         \left(\frac{\theta_I}{7.5\times10^{-4}}\right)^2 \left(\frac{f_a}{\SI{1.4e17}{\giga\electronvolt}}\right)^\frac{3}{2},\\
         \left(\frac{\theta_I}{1}\right)^2  \left(\frac{f_a}{\SI{8.7e11}{\giga\electronvolt}}\right)^\frac{7}{6},
    \end{cases}
\end{align}
where the first (second) line apply for decay constants larger (smaller)  than $\SI{1.24e17}{\giga\electronvolt}$. 
For values of $\theta_I \simeq \pi$, as needed in the region $f_a\ll 10^{12}\;\text{GeV}$, the anharmonicity \cite{Turner:1985si,Lyth:1991ub} of the potential becomes important and one can include this by multiplying the energy density with $\text{ln}\left(e/(1-\theta_I^2/\pi^2)\right)^{7/6}$\cite{Visinelli:2009zm}.
Here we chose a vacuum degeneracy (domain wall number) of $N_\text{QCD}=1$ for illustration only, which can be realized in the KSVZ model \cite{PhysRevLett.43.103,SHIFMAN1980493} or variants of the DFSZ model \cite{Zhitnitsky:1980tq,DINE1981199} with flavor specific couplings \cite{RVolkas:2023jiv}. Due to the absence of topological defects in the  preinflationary PQ breaking scenario as well as in our setup, there is no obstruction to choose $N_\text{QCD}>1$, and in all previous relations one simply has to replace $f_a$ with $f_a/N_\text{QCD}$.

When the Universe enters  radiation domination ($\omega =1/3$) one would naively expect that $R=0$. However quantum effects induce an nonzero value for $1-3 \omega$  due to the trace anomaly of non-Abelian gauge theories \cite{Kajantie:2002wa,Davoudiasl:2004gf} and we obtain $M_a(T) = c_a(T) H(T)$ in terms of the Hubble rate during radiation domination with\footnote{This expression is a factor of two larger than the result in Ref. \cite{Takahashi:2015waa} (up to a factor of  $M_\text{Pl.}/f_a$), because we consider a term $\sigma^2$ leading to $\cos(2\theta)$ instead of their $\cos(\theta)$.}
\begin{align}\label{eq:adiab}
 c_a(T) = \frac{18}{\pi} \sqrt{\frac{6}{19}} \alpha_S(T) \sqrt{ \left|\xi_\sigma\right| \cos{\left(\theta_R\right)}}.
\end{align}
Here we limit ourselves to  QCD with three flavors and three colors implying $1- 3 \omega \simeq 162 \alpha_S(T)^2/(19 \pi^2)$ \cite{Takahashi:2015waa}, where $\alpha_S$ denotes the strong coupling constant. The energy stored in the scalar fields should be subdominant to the radiation bath,
which for the axion imposes $c_a(T)< \sqrt{3/(4\pi)}M_\text{Pl.}/f_a$ and for large temperatures this is always satisfied (see Eq.~\eqref{eq:ca}), especially for $f_a\ll M_\text{Pl.}$. 

Oscillating scalar fields with a Hubble-dependent contribution to their potential, as in Eq.~\eqref{eq:adiab},  tend to track a time-dependent, adiabatically evolving minimum, which leads to a reduction of their oscillation amplitude and hence their energy density, as can be seen in the left panel of Fig.~\ref{fig:osc-1}. This \enquote{adiabatic suppression} mechanism was discovered in Ref.~\cite{Linde:1996cx} and for $c_a\gg 1$ it leads to an exponential reduction of the axion relic abundance $\Omega_a$
\begin{align}\label{eq:Omeg}
    \frac{\Omega_a (c_a\gg 1)}{\Omega_a (c_a=0)} \simeq c_a^{3p +1} e^{- \pi p c_a},
\end{align}
where $p=1/2$ for radiation domination. 
A more accurate analytical formula incorporating the time dependence of the QCD axion potential from Eq.~\eqref{eq:maT} can  be found in Ref.~\cite{Kawasaki:2017xwt}.
In our case  $c_a(T)\sim \alpha_S(T)$ is time dependent and we extract the running of $\alpha_S(T)$ for temperatures below $\SI{5}{\giga\electronvolt}$  from Ref.~\cite{Deur:2022msf}, where $\alpha_S(T\leq \SI{0.1}{\giga\electronvolt})$ freezes at $\pi$.\footnote{It is worth emphasizing that the  perturbative treatment for $c_a$ in terms of $\alpha_S$ is expected to break down around the QCD confinement scale. Reference \cite{Caldwell:2013mox} used lattice QCD data to treat the regime $100\;\text{MeV}<T<1\;\text{GeV}$ and their result imply an effective axion mass that is about an order of magnitude smaller then the number used in our analysis. Hence we treat our calculation as a conservative estimate and expect the impact of the adiabatic suppression to be even more reduced compared to our results.}
For the range above $\SI{5}{\giga\electronvolt}$ we employ the summary of measurements compiled in Ref.~\cite{Huston:2023ofk} with $\alpha_S(m_Z^2)=0.1180\pm0.0009$. The Hubble-dependent mass is bounded from above as
\begin{align}\label{eq:ca}
    c_a (T) \lesssim 10 \left(\frac{\alpha_S}{\pi}\right)  \sqrt{ \left|\xi_\sigma\right| \cos{\left(\theta_R\right)}},
\end{align}
where we used the largest possible value of  $\alpha_S(T)$. This indicates that the suppression will be less severe than expected from Eq.~\eqref{eq:Omeg}, unlike the setup considered in the Appendix of Ref.~\cite{Takahashi:2015waa}.

Oscillations starting close to the hilltop $|\theta_I|\simeq\pi$, not realizable in our case, can also counteract the aforementioned suppression to some degree, because the effective mass will not necessarily change adiabatically anymore \cite{Nakagawa:2020zjr}. It would be interesting to study the interplay of an initial velocity for the axion field $\dot{\theta}_i\neq 0$ \cite{Co:2019jts} with the adiabatic relaxation  mechanism. 
We numerically  solve the equations of motion to determine the impact of the coupling in \eqref{eq:central} on the parameter space of the postinflationary misalignment mechanism. Fitting functions for the number of relativistic degrees of freedom in energy and entropy were obtained from Ref.~\cite{Wantz:2009it}.

The impact on the parameter space reproducing the dark matter relic abundance was summarized on the right-hand side of Fig.~\ref{fig:isoc-1}.
For the largest possible value of $|\xi_\sigma|=1$ the dilution effect is most important for about  $f_a> \SI{3e11}{\giga\electronvolt}$, as the Hubble dependent mass can be much larger than the axion mass $m_a(T)$ at the temperature $T_\text{osc.}$, defined via $m_a(T_\text{osc.}) \simeq 3 H(T_\text{osc.})$, when the oscillations would usually start for $M_a=0$, which can be observed in the right panel of Fig.~\ref{fig:osc-1}.
For instance we find that $f_a\simeq 10^{16}\;\text{GeV}$, which usually requires very small values of the postinflationary misalignment angle $|\theta_I|\simeq 8.5\times10^{-3}$, known as the \enquote{anthropic window} \cite{Linde:1987bx,Wilczek:2004cr,Tegmark:2005dy,Hertzberg:2008wr}, can now work  with  $|\xi_\sigma|=1$ for the slightly larger $|\theta_I|=|\theta_R|/2\simeq 1.7\times 10^{-2}$.  
Values of $f_a \ll \SI{3e11}{\giga\electronvolt}$ are far less affected by the Hubble-dependent mass, but as usual they would only reproduce the relic abundance in the  anharmonic regime $|\theta_I|\simeq \pi$, which is inaccessible, since this quadratic model is limited to $|\theta_I|=|\theta_R|/2 \leq \pi/4$.

\textbf{Axion dark matter from fluctuations.\textemdash}
Axion dark matter from inflationary fluctuations and its impact on isocurvature modes on CMB scales was investigated in Refs.~\cite{Alonso-Alvarez:2018tus,Alonso-Alvarez:2019ixv}. With our range of $|\xi_\sigma|> 3/64$ the authors obtain the relic abundance of scalar DM in agreement with CMB isocurvature limits for e.g. $H_I \simeq 10^{13}\;\text{GeV}$ and $m_\text{DM} \gtrsim \SI{1}{\kilo\electronvolt}$. The required DM mass increases for smaller $H_I$ or larger $|\xi_\sigma|$. Realizing this mass range for the QCD axion would require  $f_a \lesssim 1~\text{TeV}$, which corresponds to a visible axion and is robustly excluded by e.g. stellar cooling bounds.

We assume that the change in the Hubble rate between inflation and the onset of reheating is adiabatic.
If that is not the case, fluctuations of the axion with a size $\delta \theta \simeq (0.01/(4|\xi_\sigma| \cos(\theta_R)))^{1/4} H_e/f_a$ \cite{Herranen:2015ima} are generated, where $H_e$ is the Hubble rate at the transition from inflation to reheating. Note that this result was derived under the assumption $|\xi_\sigma| \cos(\theta_R)\gg 1$, and if present we expect the fluctuations to be small due to the assumption of preinflationary PQ breaking in Eq.~\eqref{eq:max} and the fact that $H_e<H_I$. Numerically we find for a GUT scale axion  and potentially observable tensor modes  $H_I=10^{13}\;\text{GeV}$ (assuming $H_e\simeq H_I$), which requires $\theta_I=1.7\times10^{-2}$ to reproduce the dark matter abundance,  that
\begin{align}
    \delta \theta \simeq \frac{2.2\times10^{-4}}{(|\xi_\sigma| \cos(\theta_R))^\frac{1}{4}} \left(\frac{H_e}{10^{13}\;\text{GeV}}\right) \left(\frac{10^{16}\;\text{GeV}}{f_a}\right),
\end{align}
which is sufficiently small compared to $\theta_I$ for our range of $3/64 < |\xi_\sigma| \cos(\theta_R) < 1$. These fluctuations will not affect the CMB as they are generated at the very end of inflation, 50-60 e-folds after the scales that get imprinted on the CMB have left the horizon.

During (p)reheating with an inflaton oscillating in a potential, that can be approximated as quadratic, the Ricci scalar will oscillate too. Hence the effective axion mass will oscillate and become temporarily tachyonic. Such a behavior can lead to production of axion fluctuations via the geometric preheating scenario \cite{Bassett:1997az}. References \cite{Dufaux:2006ee,Markkanen:2015xuw} used this effect to generate the dark matter relic abundance and ref.~\cite{Alonso-Alvarez:2018tus} found that it is negligible for $m_\text{DM} \lesssim \SI{10}{\giga\electronvolt}$ or $|\xi_\sigma| \leq 1$, which are both satisfied for our setup. In general this will come into effect, when the adiabaticity condition  {$|\dot{M_a}/M_a^2| < 1$ is violated. Assuming that the inflaton and hence the Ricci scalar oscillate with a timescale of $t_\text{RH}= 1/(3 H_\text{RH})$ in terms of the Hubble rate during (p)reheating $H_\text{RH}$ (see Ref.~\cite{Chung:2018ayg} for the case of faster oscillations), we find  $\dot{M_a}\simeq 3 M_a  H_\text{RH}$ and the adiabaticity condition implies
\begin{align}\label{eq:fluct}
    |\xi_\sigma| \cos(\theta_R) > \frac{3}{16},
\end{align}
which is roughly a factor of two larger than the previous lower bound in Eq.~\eqref{eq:xi}. In general, if the duration of the reheating epoch is much shorter than the timescale over which $M_a^2$ turns tachyonic, we do not have to worry about the excursion of the axion zero mode and retain the relation $\theta_I =\theta_R / 2$.
Thus the evolution of the axion during reheating is fairly model dependent.
One viable scenario could be instantaneous reheating from  the perturbative decay of the inflaton. Warm inflation \cite{Bartrum:2013fia}, where  the slow-roll phase is directly followed by radiation domination, features neither an epoch of reheating nor an oscillating inflaton. 

 During radiation domination we find that the  evolution of the adiabaticity parameter $|\dot{M_a}/M_a^2|$ depends on the interplay of the Hubble rate, which decreases as the temperature falls, and the strong gauge coupling appearing in Eq.~\eqref{eq:adiab}, that increases. Due to these different behaviors  we find by numerical extrapolation  that the adiabaticity parameter first grows and typically crosses unity at at temperatures of  $T_*=\mathcal{O}(\SI{1}{\tera\electronvolt})$, before reaching a maximum  around $\SI{300}{\giga\electronvolt}$, and subsequently decreasing below unity. The maximum of $|\dot{M_a}/M_a^2|$ stays below unity as long as $|\xi_\sigma| \cos(\theta_R) \gtrsim 25$, which  can be made compatible with the upper bound in Eq.~\eqref{eq:xi} for $m_\rho \gg 35 H_I$. Even when the adiabaticity condition is violated at $T_*$ we do not expect a significant contribution to the axion relic density (see e.g. \cite{Babichev:2020xeg}), as the resulting energy density will scale with powers of the Hubble rate at $T_*$. Since this Hubble rate is orders of magnitude smaller than the Hubble rate during inflation or (p)reheating, the quantum fluctuations during radiation domination will be even more suppressed compared to those already subleading contributions discussed earlier in this section.

We conclude that fluctuations either during inflation,  (p)reheating or radiation domination are negligible for the axion yield, and hence, the later vacuum realignment discussed in the previous section dominates the relic density.

\textbf{Summary.\textemdash}We proposed to solve the axion isocurvature problem with a nonminimal coupling of the PQ breaking gauge singlet scalar to gravity. Such a coupling explicitly breaks PQ symmetry during inflation and thus lifts the axion mass above the inflationary Hubble scale. This modification of the axion potential is negligible at present times due to the smallness of today's Hubble rate. GUT-scale PQ breaking ($f_a = 10^{16}\;\text{GeV}$) is compatible with a large Hubble rate during inflation. We investigated the impact of the adiabatically relaxing minimum as well as the post-inflationary misalignment angle from the phase of the nonminimal coupling  on the axion oscillations, that here explain the dark matter abundance for about $f_a> \SI{3e11}{\giga\electronvolt}$. The smoking gun signals for singling out the proposed scenario are the simultaneous detection of inflationary gravitational waves at next generation CMB experiments such as CMB-S4 or LiteBIRD together with the detection of a GUT-scale axion at the haloscopes ABRACADABRA, DMRadio or CASPEr-Electric. Our study motivates further investigations of $CP$-violation from quantum gravity.

\textbf{Acknowledgements.\textemdash}We would like to thank Giacomo Landini, Peera Simakachorn and Andreas Trautner for valuable feedback on the manuscript, as well as Maya Hager for discussions about the value of the  Ricci scalar in the solar system.
MB is supported by \enquote{Consolidación Investigadora Grant CNS2022-135592}, funded also by \enquote{European Union NextGenerationEU/PRTR}.

\bibliographystyle{utphys}
\bibliography{referencesIsoc}

\providecommand{\href}[2]{#2}\begingroup\raggedright\begin{thebibliography}{100}

\bibitem{PhysRevLett.38.1440}
R.~D. Peccei and H.~R. Quinn, ``$\mathrm{CP}$ conservation in the presence of
  pseudoparticles,'' \href{http://dx.doi.org/10.1103/PhysRevLett.38.1440}{{\em
  Phys. Rev. Lett.} {\bfseries 38} (Jun, 1977) 1440--1443}.
  \url{https://link.aps.org/doi/10.1103/PhysRevLett.38.1440}.

\bibitem{Peccei:1977ur}
R.~D. Peccei and H.~R. Quinn, ``{Constraints Imposed by CP Conservation in the
  Presence of Instantons},''
  \href{http://dx.doi.org/10.1103/PhysRevD.16.1791}{{\em Phys. Rev. D}
  {\bfseries 16} (1977) 1791--1797}.

\bibitem{PhysRevLett.40.223}
S.~Weinberg, ``A new light boson?,''
  \href{http://dx.doi.org/10.1103/PhysRevLett.40.223}{{\em Phys. Rev. Lett.}
  {\bfseries 40} (Jan, 1978) 223--226}.
  \url{https://link.aps.org/doi/10.1103/PhysRevLett.40.223}.

\bibitem{PhysRevLett.40.279}
F.~Wilczek, ``Problem of strong $p$ and $t$ invariance in the presence of
  instantons,'' \href{http://dx.doi.org/10.1103/PhysRevLett.40.279}{{\em Phys.
  Rev. Lett.} {\bfseries 40} (Jan, 1978) 279--282}.
  \url{https://link.aps.org/doi/10.1103/PhysRevLett.40.279}.

\bibitem{DiLuzio:2020wdo}
L.~Di~Luzio, M.~Giannotti, E.~Nardi, and L.~Visinelli, ``{The landscape of QCD
  axion models},'' \href{http://dx.doi.org/10.1016/j.physrep.2020.06.002}{{\em
  Phys. Rept.} {\bfseries 870} (2020) 1--117},
  \href{http://arxiv.org/abs/2003.01100}{{\ttfamily arXiv:2003.01100
  [hep-ph]}}.

\bibitem{OHare:2024nmr}
C.~A.~J. O'Hare, ``{Cosmology of axion dark matter},''
  \href{http://arxiv.org/abs/2403.17697}{{\ttfamily arXiv:2403.17697
  [hep-ph]}}.

\bibitem{Abbott:1982af}
L.~F. Abbott and P.~Sikivie, ``{A Cosmological Bound on the Invisible Axion},''
  \href{http://dx.doi.org/10.1016/0370-2693(83)90638-X}{{\em Phys. Lett. B}
  {\bfseries 120} (1983) 133--136}.

\bibitem{Dine:1982ah}
M.~Dine and W.~Fischler, ``{The Not So Harmless Axion},''
  \href{http://dx.doi.org/10.1016/0370-2693(83)90639-1}{{\em Phys. Lett. B}
  {\bfseries 120} (1983) 137--141}.

\bibitem{Preskill:1982cy}
J.~Preskill, M.~B. Wise, and F.~Wilczek, ``{Cosmology of the Invisible
  Axion},'' \href{http://dx.doi.org/10.1016/0370-2693(83)90637-8}{{\em Phys.
  Lett. B} {\bfseries 120} (1983) 127--132}.

\bibitem{Redi:2022llj}
M.~Redi and A.~Tesi, ``{Meso-inflationary QCD axion},''
  \href{http://dx.doi.org/10.1103/PhysRevD.107.095032}{{\em Phys. Rev. D}
  {\bfseries 107} no.~9, (2023) 095032},
  \href{http://arxiv.org/abs/2211.06421}{{\ttfamily arXiv:2211.06421
  [hep-ph]}}.

\bibitem{Gorghetto:2023vqu}
M.~Gorghetto, E.~Hardy, H.~Nicolaescu, A.~Notari, and M.~Redi, ``{Early vs late
  string networks from a minimal QCD Axion},''
  \href{http://dx.doi.org/10.1007/JHEP02(2024)223}{{\em JHEP} {\bfseries 02}
  (2024) 223}, \href{http://arxiv.org/abs/2311.09315}{{\ttfamily
  arXiv:2311.09315 [hep-ph]}}.

\bibitem{PhysRevD152738}
G.~W. Gibbons and S.~W. Hawking, ``Cosmological event horizons, thermodynamics,
  and particle creation,''
  \href{http://dx.doi.org/10.1103/PhysRevD.15.2738}{{\em Phys. Rev. D}
  {\bfseries 15} (May, 1977) 2738--2751}.
  \url{https://link.aps.org/doi/10.1103/PhysRevD.15.2738}.

\bibitem{Giudice:2000ex}
G.~F. Giudice, E.~W. Kolb, and A.~Riotto, ``{Largest temperature of the
  radiation era and its cosmological implications},''
  \href{http://dx.doi.org/10.1103/PhysRevD.64.023508}{{\em Phys. Rev. D}
  {\bfseries 64} (2001) 023508},
  \href{http://arxiv.org/abs/hep-ph/0005123}{{\ttfamily arXiv:hep-ph/0005123}}.

\bibitem{Kolb:2003ke}
E.~W. Kolb, A.~Notari, and A.~Riotto, ``{On the reheating stage after
  inflation},'' \href{http://dx.doi.org/10.1103/PhysRevD.68.123505}{{\em Phys.
  Rev. D} {\bfseries 68} (2003) 123505},
  \href{http://arxiv.org/abs/hep-ph/0307241}{{\ttfamily arXiv:hep-ph/0307241}}.

\bibitem{Co:2020xaf}
R.~T. Co, E.~Gonzalez, and K.~Harigaya, ``{Increasing Temperature toward the
  Completion of Reheating},''
  \href{http://dx.doi.org/10.1088/1475-7516/2020/11/038}{{\em JCAP} {\bfseries
  11} (2020) 038}, \href{http://arxiv.org/abs/2007.04328}{{\ttfamily
  arXiv:2007.04328 [astro-ph.CO]}}.

\bibitem{Planck:2018jri}
{\bfseries Planck} Collaboration, Y.~Akrami {\em et~al.}, ``{Planck 2018
  results. X. Constraints on inflation},''
  \href{http://dx.doi.org/10.1051/0004-6361/201833887}{{\em Astron. Astrophys.}
  {\bfseries 641} (2020) A10},
  \href{http://arxiv.org/abs/1807.06211}{{\ttfamily arXiv:1807.06211
  [astro-ph.CO]}}.

\bibitem{Hamann:2009yf}
J.~Hamann, S.~Hannestad, G.~G. Raffelt, and Y.~Y.~Y. Wong, ``{Isocurvature
  forecast in the anthropic axion window},''
  \href{http://dx.doi.org/10.1088/1475-7516/2009/06/022}{{\em JCAP} {\bfseries
  06} (2009) 022}, \href{http://arxiv.org/abs/0904.0647}{{\ttfamily
  arXiv:0904.0647 [hep-ph]}}.

\bibitem{Turner:1985si}
M.~S. Turner, ``{Cosmic and Local Mass Density of Invisible Axions},''
  \href{http://dx.doi.org/10.1103/PhysRevD.33.889}{{\em Phys. Rev. D}
  {\bfseries 33} (1986) 889--896}.

\bibitem{Lyth:1991ub}
D.~H. Lyth, ``{Axions and inflation: Sitting in the vacuum},''
  \href{http://dx.doi.org/10.1103/PhysRevD.45.3394}{{\em Phys. Rev. D}
  {\bfseries 45} (1992) 3394--3404}.

\bibitem{Visinelli:2009zm}
L.~Visinelli and P.~Gondolo, ``{Dark Matter Axions Revisited},''
  \href{http://dx.doi.org/10.1103/PhysRevD.80.035024}{{\em Phys. Rev. D}
  {\bfseries 80} (2009) 035024},
  \href{http://arxiv.org/abs/0903.4377}{{\ttfamily arXiv:0903.4377
  [astro-ph.CO]}}.

\bibitem{Kobayashi:2013nva}
T.~Kobayashi, R.~Kurematsu, and F.~Takahashi, ``{Isocurvature Constraints and
  Anharmonic Effects on QCD Axion Dark Matter},''
  \href{http://dx.doi.org/10.1088/1475-7516/2013/09/032}{{\em JCAP} {\bfseries
  09} (2013) 032}, \href{http://arxiv.org/abs/1304.0922}{{\ttfamily
  arXiv:1304.0922 [hep-ph]}}.

\bibitem{GrillidiCortona:2015jxo}
G.~Grilli~di Cortona, E.~Hardy, J.~Pardo~Vega, and G.~Villadoro, ``{The QCD
  axion, precisely},'' \href{http://dx.doi.org/10.1007/JHEP01(2016)034}{{\em
  JHEP} {\bfseries 01} (2016) 034},
  \href{http://arxiv.org/abs/1511.02867}{{\ttfamily arXiv:1511.02867
  [hep-ph]}}.

\bibitem{Wise:1981ry}
M.~B. Wise, H.~Georgi, and S.~L. Glashow, ``{SU(5) and the Invisible Axion},''
  \href{http://dx.doi.org/10.1103/PhysRevLett.47.402}{{\em Phys. Rev. Lett.}
  {\bfseries 47} (1981) 402}.

\bibitem{Lazarides:1981kz}
G.~Lazarides, ``{SO(10) and the Invisible Axion},''
  \href{http://dx.doi.org/10.1103/PhysRevD.25.2425}{{\em Phys. Rev. D}
  {\bfseries 25} (1982) 2425}.

\bibitem{Choi:1985je}
K.~Choi and J.~E. Kim, ``{Harmful Axions in Superstring Models},''
  \href{http://dx.doi.org/10.1016/0370-2693(85)90416-2}{{\em Phys. Lett. B}
  {\bfseries 154} (1985) 393}. [Erratum: Phys.Lett.B 156, 452 (1985)].

\bibitem{Svrcek:2006yi}
P.~Svrcek and E.~Witten, ``{Axions In String Theory},''
  \href{http://dx.doi.org/10.1088/1126-6708/2006/06/051}{{\em JHEP} {\bfseries
  06} (2006) 051}, \href{http://arxiv.org/abs/hep-th/0605206}{{\ttfamily
  arXiv:hep-th/0605206}}.

\bibitem{Arvanitaki:2009fg}
A.~Arvanitaki, S.~Dimopoulos, S.~Dubovsky, N.~Kaloper, and J.~March-Russell,
  ``{String Axiverse},''
  \href{http://dx.doi.org/10.1103/PhysRevD.81.123530}{{\em Phys. Rev. D}
  {\bfseries 81} (2010) 123530},
  \href{http://arxiv.org/abs/0905.4720}{{\ttfamily arXiv:0905.4720 [hep-th]}}.

\bibitem{Kahn:2016aff}
Y.~Kahn, B.~R. Safdi, and J.~Thaler, ``{Broadband and Resonant Approaches to
  Axion Dark Matter Detection},''
  \href{http://dx.doi.org/10.1103/PhysRevLett.117.141801}{{\em Phys. Rev.
  Lett.} {\bfseries 117} no.~14, (2016) 141801},
  \href{http://arxiv.org/abs/1602.01086}{{\ttfamily arXiv:1602.01086
  [hep-ph]}}.

\bibitem{DMRadio:2022pkf}
{\bfseries DMRadio} Collaboration, L.~Brouwer {\em et~al.}, ``{Projected
  sensitivity of DMRadio-m3: A search for the QCD axion below
  1\,\,\ensuremath{\mu}eV},''
  \href{http://dx.doi.org/10.1103/PhysRevD.106.103008}{{\em Phys. Rev. D}
  {\bfseries 106} no.~10, (2022) 103008},
  \href{http://arxiv.org/abs/2204.13781}{{\ttfamily arXiv:2204.13781
  [hep-ex]}}.

\bibitem{DMRadio:2022jfv}
{\bfseries DMRadio} Collaboration, L.~Brouwer {\em et~al.}, ``{Proposal for a
  definitive search for GUT-scale QCD axions},''
  \href{http://dx.doi.org/10.1103/PhysRevD.106.112003}{{\em Phys. Rev. D}
  {\bfseries 106} no.~11, (2022) 112003},
  \href{http://arxiv.org/abs/2203.11246}{{\ttfamily arXiv:2203.11246
  [hep-ex]}}.

\bibitem{Graham:2013gfa}
P.~W. Graham and S.~Rajendran, ``{New Observables for Direct Detection of Axion
  Dark Matter},'' \href{http://dx.doi.org/10.1103/PhysRevD.88.035023}{{\em
  Phys. Rev. D} {\bfseries 88} (2013) 035023},
  \href{http://arxiv.org/abs/1306.6088}{{\ttfamily arXiv:1306.6088 [hep-ph]}}.

\bibitem{Budker:2013hfa}
D.~Budker, P.~W. Graham, M.~Ledbetter, S.~Rajendran, and A.~Sushkov,
  ``{Proposal for a Cosmic Axion Spin Precession Experiment (CASPEr)},''
  \href{http://dx.doi.org/10.1103/PhysRevX.4.021030}{{\em Phys. Rev. X}
  {\bfseries 4} no.~2, (2014) 021030},
  \href{http://arxiv.org/abs/1306.6089}{{\ttfamily arXiv:1306.6089 [hep-ph]}}.

\bibitem{JacksonKimball:2017elr}
D.~F. Jackson~Kimball {\em et~al.}, ``{Overview of the Cosmic Axion Spin
  Precession Experiment (CASPEr)},''
  \href{http://dx.doi.org/10.1007/978-3-030-43761-9_13}{{\em Springer Proc.
  Phys.} {\bfseries 245} (2020) 105--121},
  \href{http://arxiv.org/abs/1711.08999}{{\ttfamily arXiv:1711.08999
  [physics.ins-det]}}.

\bibitem{CMB-S4:2016ple}
{\bfseries CMB-S4} Collaboration, K.~N. Abazajian {\em et~al.}, ``{CMB-S4
  Science Book, First Edition},''
  \href{http://arxiv.org/abs/1610.02743}{{\ttfamily arXiv:1610.02743
  [astro-ph.CO]}}.

\bibitem{Matsumura:2013aja}
T.~Matsumura {\em et~al.}, ``{Mission design of LiteBIRD},''
  \href{http://dx.doi.org/10.1007/s10909-013-0996-1}{{\em J. Low Temp. Phys.}
  {\bfseries 176} (2014) 733}, \href{http://arxiv.org/abs/1311.2847}{{\ttfamily
  arXiv:1311.2847 [astro-ph.IM]}}.

\bibitem{BICEP:2021xfz}
{\bfseries BICEP, Keck} Collaboration, P.~A.~R. Ade {\em et~al.}, ``{Improved
  Constraints on Primordial Gravitational Waves using Planck, WMAP, and
  BICEP/Keck Observations through the 2018 Observing Season},''
  \href{http://dx.doi.org/10.1103/PhysRevLett.127.151301}{{\em Phys. Rev.
  Lett.} {\bfseries 127} no.~15, (2021) 151301},
  \href{http://arxiv.org/abs/2110.00483}{{\ttfamily arXiv:2110.00483
  [astro-ph.CO]}}.

\bibitem{Dvali:1995ce}
G.~R. Dvali, ``{Removing the cosmological bound on the axion scale},''
  \href{http://arxiv.org/abs/hep-ph/9505253}{{\ttfamily arXiv:hep-ph/9505253}}.

\bibitem{Jeong:2013xta}
K.~S. Jeong and F.~Takahashi, ``{Suppressing Isocurvature Perturbations of QCD
  Axion Dark Matter},''
  \href{http://dx.doi.org/10.1016/j.physletb.2013.10.061}{{\em Phys. Lett. B}
  {\bfseries 727} (2013) 448--451},
  \href{http://arxiv.org/abs/1304.8131}{{\ttfamily arXiv:1304.8131 [hep-ph]}}.

\bibitem{Koutsangelas:2022lte}
E.~Koutsangelas, ``{Removing the cosmological bound on the axion scale in the
  Kim-Shifman-Vainshtein-Zakharov and Dine-Fischler-Srednicki-Zhitnitsky
  models},'' \href{http://dx.doi.org/10.1103/PhysRevD.107.095009}{{\em Phys.
  Rev. D} {\bfseries 107} no.~9, (2023) 095009},
  \href{http://arxiv.org/abs/2212.07822}{{\ttfamily arXiv:2212.07822
  [hep-ph]}}.

\bibitem{Takahashi:2015waa}
F.~Takahashi and M.~Yamada, ``{Strongly broken Peccei-Quinn symmetry in the
  early Universe},''
  \href{http://dx.doi.org/10.1088/1475-7516/2015/10/010}{{\em JCAP} {\bfseries
  10} (2015) 010}, \href{http://arxiv.org/abs/1507.06387}{{\ttfamily
  arXiv:1507.06387 [hep-ph]}}.

\bibitem{Nomura:2015xil}
Y.~Nomura, S.~Rajendran, and F.~Sanches, ``{Axion Isocurvature and Magnetic
  Monopoles},'' \href{http://dx.doi.org/10.1103/PhysRevLett.116.141803}{{\em
  Phys. Rev. Lett.} {\bfseries 116} no.~14, (2016) 141803},
  \href{http://arxiv.org/abs/1511.06347}{{\ttfamily arXiv:1511.06347
  [hep-ph]}}.

\bibitem{Kawasaki:2017xwt}
M.~Kawasaki, F.~Takahashi, and M.~Yamada, ``{Adiabatic suppression of the axion
  abundance and isocurvature due to coupling to hidden monopoles},''
  \href{http://dx.doi.org/10.1007/JHEP01(2018)053}{{\em JHEP} {\bfseries 01}
  (2018) 053}, \href{http://arxiv.org/abs/1708.06047}{{\ttfamily
  arXiv:1708.06047 [hep-ph]}}.

\bibitem{Witten:1979ey}
E.~Witten, ``{Dyons of Charge e theta/2 pi},''
  \href{http://dx.doi.org/10.1016/0370-2693(79)90838-4}{{\em Phys. Lett. B}
  {\bfseries 86} (1979) 283--287}.

\bibitem{Folkerts:2013tua}
S.~Folkerts, C.~Germani, and J.~Redondo, ``{Axion Dark Matter and Planck favor
  non-minimal couplings to gravity},''
  \href{http://dx.doi.org/10.1016/j.physletb.2013.12.026}{{\em Phys. Lett. B}
  {\bfseries 728} (2014) 532--536},
  \href{http://arxiv.org/abs/1304.7270}{{\ttfamily arXiv:1304.7270 [hep-ph]}}.

\bibitem{Linde:1991km}
A.~D. Linde, ``{Axions in inflationary cosmology},''
  \href{http://dx.doi.org/10.1016/0370-2693(91)90130-I}{{\em Phys. Lett. B}
  {\bfseries 259} (1991) 38--47}.

\bibitem{Choi:2014uaa}
K.~Choi, K.~S. Jeong, and M.-S. Seo, ``{String theoretic QCD axions in the
  light of PLANCK and BICEP2},''
  \href{http://dx.doi.org/10.1007/JHEP07(2014)092}{{\em JHEP} {\bfseries 07}
  (2014) 092}, \href{http://arxiv.org/abs/1404.3880}{{\ttfamily arXiv:1404.3880
  [hep-th]}}.

\bibitem{Chun:2014xva}
E.~J. Chun, ``{Axion Dark Matter with High-Scale Inflation},''
  \href{http://dx.doi.org/10.1016/j.physletb.2014.06.017}{{\em Phys. Lett. B}
  {\bfseries 735} (2014) 164--167},
  \href{http://arxiv.org/abs/1404.4284}{{\ttfamily arXiv:1404.4284 [hep-ph]}}.

\bibitem{Higaki:2014ooa}
T.~Higaki, K.~S. Jeong, and F.~Takahashi, ``{Solving the Tension between
  High-Scale Inflation and Axion Isocurvature Perturbations},''
  \href{http://dx.doi.org/10.1016/j.physletb.2014.05.014}{{\em Phys. Lett. B}
  {\bfseries 734} (2014) 21--26},
  \href{http://arxiv.org/abs/1403.4186}{{\ttfamily arXiv:1403.4186 [hep-ph]}}.

\bibitem{Fairbairn:2014zta}
M.~Fairbairn, R.~Hogan, and D.~J.~E. Marsh, ``{Unifying inflation and dark
  matter with the Peccei-Quinn field: observable axions and observable
  tensors},'' \href{http://dx.doi.org/10.1103/PhysRevD.91.023509}{{\em Phys.
  Rev. D} {\bfseries 91} no.~2, (2015) 023509},
  \href{http://arxiv.org/abs/1410.1752}{{\ttfamily arXiv:1410.1752 [hep-ph]}}.

\bibitem{Nakayama:2015pba}
K.~Nakayama and M.~Takimoto, ``{Higgs inflation and suppression of axion
  isocurvature perturbation},''
  \href{http://dx.doi.org/10.1016/j.physletb.2015.07.001}{{\em Phys. Lett. B}
  {\bfseries 748} (2015) 108--112},
  \href{http://arxiv.org/abs/1505.02119}{{\ttfamily arXiv:1505.02119
  [hep-ph]}}.

\bibitem{Harigaya:2015hha}
K.~Harigaya, M.~Ibe, M.~Kawasaki, and T.~T. Yanagida, ``{Dynamics of
  Peccei-Quinn Breaking Field after Inflation and Axion Isocurvature
  Perturbations},'' \href{http://dx.doi.org/10.1088/1475-7516/2015/11/003}{{\em
  JCAP} {\bfseries 11} (2015) 003},
  \href{http://arxiv.org/abs/1507.00119}{{\ttfamily arXiv:1507.00119
  [hep-ph]}}.

\bibitem{Kearney:2016vqw}
J.~Kearney, N.~Orlofsky, and A.~Pierce, ``{High-Scale Axions without
  Isocurvature from Inflationary Dynamics},''
  \href{http://dx.doi.org/10.1103/PhysRevD.93.095026}{{\em Phys. Rev. D}
  {\bfseries 93} no.~9, (2016) 095026},
  \href{http://arxiv.org/abs/1601.03049}{{\ttfamily arXiv:1601.03049
  [hep-ph]}}.

\bibitem{Bao:2022hsg}
Y.~Bao, J.~Fan, and L.~Li, ``{Opening up a Window on the Postinflationary QCD
  Axion},'' \href{http://dx.doi.org/10.1103/PhysRevLett.130.241001}{{\em Phys.
  Rev. Lett.} {\bfseries 130} no.~24, (2023) 241001},
  \href{http://arxiv.org/abs/2209.09908}{{\ttfamily arXiv:2209.09908
  [hep-ph]}}.

\bibitem{Kawasaki:2017ycl}
M.~Kawasaki, L.~Pearce, L.~Yang, and A.~Kusenko, ``{Relaxation leptogenesis,
  isocurvature perturbations, and the cosmic infrared background},''
  \href{http://dx.doi.org/10.1103/PhysRevD.95.103006}{{\em Phys. Rev. D}
  {\bfseries 95} no.~10, (2017) 103006},
  \href{http://arxiv.org/abs/1701.02175}{{\ttfamily arXiv:1701.02175
  [hep-ph]}}.

\bibitem{Rosa:2021gbe}
J.~a.~G. Rosa and L.~B. Ventura, ``{Spontaneous breaking of the Peccei-Quinn
  symmetry during warm inflation},''
  \href{http://arxiv.org/abs/2105.05771}{{\ttfamily arXiv:2105.05771
  [hep-ph]}}.

\bibitem{Cardoso:2018tly}
V.~Cardoso, O.~J.~C. Dias, G.~S. Hartnett, M.~Middleton, P.~Pani, and J.~E.
  Santos, ``{Constraining the mass of dark photons and axion-like particles
  through black-hole superradiance},''
  \href{http://dx.doi.org/10.1088/1475-7516/2018/03/043}{{\em JCAP} {\bfseries
  03} (2018) 043}, \href{http://arxiv.org/abs/1801.01420}{{\ttfamily
  arXiv:1801.01420 [gr-qc]}}.

\bibitem{Fischer:2016cyd}
T.~Fischer, S.~Chakraborty, M.~Giannotti, A.~Mirizzi, A.~Payez, and
  A.~Ringwald, ``{Probing axions with the neutrino signal from the next
  galactic supernova},''
  \href{http://dx.doi.org/10.1103/PhysRevD.94.085012}{{\em Phys. Rev. D}
  {\bfseries 94} no.~8, (2016) 085012},
  \href{http://arxiv.org/abs/1605.08780}{{\ttfamily arXiv:1605.08780
  [astro-ph.HE]}}.

\bibitem{Carenza:2019pxu}
P.~Carenza, T.~Fischer, M.~Giannotti, G.~Guo, G.~Mart\'\i{}nez-Pinedo, and
  A.~Mirizzi, ``{Improved axion emissivity from a supernova via nucleon-nucleon
  bremsstrahlung},''
  \href{http://dx.doi.org/10.1088/1475-7516/2019/10/016}{{\em JCAP} {\bfseries
  10} no.~10, (2019) 016}, \href{http://arxiv.org/abs/1906.11844}{{\ttfamily
  arXiv:1906.11844 [hep-ph]}}. [Erratum: JCAP 05, E01 (2020)].

\bibitem{Hashimoto:2021xgu}
T.~Hashimoto, N.~S. Risdianto, and D.~Suematsu, ``{Inflation connected to the
  origin of CP violation},''
  \href{http://dx.doi.org/10.1103/PhysRevD.104.075034}{{\em Phys. Rev. D}
  {\bfseries 104} no.~7, (2021) 075034},
  \href{http://arxiv.org/abs/2105.06089}{{\ttfamily arXiv:2105.06089
  [hep-ph]}}.

\bibitem{Barenboim:2024akt}
G.~Barenboim, P.~Ko, and W.-i. Park, ``{The minimal cosmological standard
  model},'' \href{http://arxiv.org/abs/2403.05390}{{\ttfamily arXiv:2403.05390
  [hep-ph]}}.

\bibitem{Barenboim:2024xxa}
G.~Barenboim, P.~Ko, and W.-i. Park, ``{Axi-majoron for almost everything},''
  \href{http://arxiv.org/abs/2403.08675}{{\ttfamily arXiv:2403.08675
  [hep-ph]}}.

\bibitem{Himmetoglu:2008zp}
B.~Himmetoglu, C.~R. Contaldi, and M.~Peloso, ``{Instability of anisotropic
  cosmological solutions supported by vector fields},''
  \href{http://dx.doi.org/10.1103/PhysRevLett.102.111301}{{\em Phys. Rev.
  Lett.} {\bfseries 102} (2009) 111301},
  \href{http://arxiv.org/abs/0809.2779}{{\ttfamily arXiv:0809.2779
  [astro-ph]}}.

\bibitem{Himmetoglu:2009qi}
B.~Himmetoglu, C.~R. Contaldi, and M.~Peloso, ``{Ghost instabilities of
  cosmological models with vector fields nonminimally coupled to the
  curvature},'' \href{http://dx.doi.org/10.1103/PhysRevD.80.123530}{{\em Phys.
  Rev. D} {\bfseries 80} (2009) 123530},
  \href{http://arxiv.org/abs/0909.3524}{{\ttfamily arXiv:0909.3524
  [astro-ph.CO]}}.

\bibitem{Karciauskas:2010as}
M.~Karciauskas and D.~H. Lyth, ``{On the health of a vector field with (R
  A\textasciicircum{}2)/6 coupling to gravity},''
  \href{http://dx.doi.org/10.1088/1475-7516/2010/11/023}{{\em JCAP} {\bfseries
  11} (2010) 023}, \href{http://arxiv.org/abs/1007.1426}{{\ttfamily
  arXiv:1007.1426 [astro-ph.CO]}}.

\bibitem{Arias:2012az}
P.~Arias, D.~Cadamuro, M.~Goodsell, J.~Jaeckel, J.~Redondo, and A.~Ringwald,
  ``{WISPy Cold Dark Matter},''
  \href{http://dx.doi.org/10.1088/1475-7516/2012/06/013}{{\em JCAP} {\bfseries
  06} (2012) 013}, \href{http://arxiv.org/abs/1201.5902}{{\ttfamily
  arXiv:1201.5902 [hep-ph]}}.

\bibitem{Alonso-Alvarez:2019ixv}
G.~Alonso-\'Alvarez, T.~Hugle, and J.~Jaeckel, ``{Misalignment
  \textbackslash{}\& Co.: (Pseudo-)scalar and vector dark matter with curvature
  couplings},'' \href{http://dx.doi.org/10.1088/1475-7516/2020/02/014}{{\em
  JCAP} {\bfseries 02} (2020) 014},
  \href{http://arxiv.org/abs/1905.09836}{{\ttfamily arXiv:1905.09836
  [hep-ph]}}.

\bibitem{Nakayama:2019rhg}
K.~Nakayama, ``{Vector Coherent Oscillation Dark Matter},''
  \href{http://dx.doi.org/10.1088/1475-7516/2019/10/019}{{\em JCAP} {\bfseries
  10} (2019) 019}, \href{http://arxiv.org/abs/1907.06243}{{\ttfamily
  arXiv:1907.06243 [hep-ph]}}.

\bibitem{Hell:2024xbv}
A.~Hell, ``{Unveiling the inconsistency of the Proca theory with non-minimal
  coupling to gravity},'' \href{http://arxiv.org/abs/2403.18673}{{\ttfamily
  arXiv:2403.18673 [gr-qc]}}.

\bibitem{Birrell:1982ix}
N.~D. Birrell and P.~C.~W. Davies,
  \href{http://dx.doi.org/10.1017/CBO9780511622632}{{\em {Quantum Fields in
  Curved Space}}}.
\newblock Cambridge Monographs on Mathematical Physics. Cambridge Univ. Press,
  Cambridge, UK, 2, 1984.

\bibitem{Parker:2009uva}
L.~E. Parker and D.~Toms,
  \href{http://dx.doi.org/10.1017/CBO9780511813924}{{\em {Quantum Field Theory
  in Curved Spacetime}: {Quantized Field and Gravity}}}.
\newblock Cambridge Monographs on Mathematical Physics. Cambridge University
  Press, 8, 2009.

\bibitem{Ballesteros:2016xej}
G.~Ballesteros, J.~Redondo, A.~Ringwald, and C.~Tamarit, ``{Standard
  Model\textemdash{}axion\textemdash{}seesaw\textemdash{}Higgs portal
  inflation. Five problems of particle physics and cosmology solved in one
  stroke},'' \href{http://dx.doi.org/10.1088/1475-7516/2017/08/001}{{\em JCAP}
  {\bfseries 08} (2017) 001}, \href{http://arxiv.org/abs/1610.01639}{{\ttfamily
  arXiv:1610.01639 [hep-ph]}}.

\bibitem{Boucenna:2017fna}
S.~M. Boucenna and Q.~Shafi, ``{Axion inflation, proton decay, and leptogenesis
  in $SU(5)\times U(1)_{PQ}$},''
  \href{http://dx.doi.org/10.1103/PhysRevD.97.075012}{{\em Phys. Rev. D}
  {\bfseries 97} no.~7, (2018) 075012},
  \href{http://arxiv.org/abs/1712.06526}{{\ttfamily arXiv:1712.06526
  [hep-ph]}}.

\bibitem{Hamaguchi:2021mmt}
K.~Hamaguchi, Y.~Kanazawa, and N.~Nagata, ``{Axion quality problem alleviated
  by nonminimal coupling to gravity},''
  \href{http://dx.doi.org/10.1103/PhysRevD.105.076008}{{\em Phys. Rev. D}
  {\bfseries 105} no.~7, (2022) 076008},
  \href{http://arxiv.org/abs/2108.13245}{{\ttfamily arXiv:2108.13245
  [hep-th]}}.

\bibitem{DalCin:2023uai}
D.~Dal~Cin and T.~Kobayashi, ``{Ultraviolet sensitivity of Peccei-Quinn
  inflation},'' \href{http://dx.doi.org/10.1103/PhysRevD.108.063530}{{\em Phys.
  Rev. D} {\bfseries 108} no.~6, (2023) 063530},
  \href{http://arxiv.org/abs/2305.18524}{{\ttfamily arXiv:2305.18524
  [hep-ph]}}.

\bibitem{McDonough:2020gmn}
E.~McDonough, A.~H. Guth, and D.~I. Kaiser, ``{Nonminimal Couplings and the
  Forgotten Field of Axion Inflation},''
  \href{http://arxiv.org/abs/2010.04179}{{\ttfamily arXiv:2010.04179
  [hep-th]}}.

\bibitem{Lerner:2009na}
R.~N. Lerner and J.~McDonald, ``{Higgs Inflation and Naturalness},''
  \href{http://dx.doi.org/10.1088/1475-7516/2010/04/015}{{\em JCAP} {\bfseries
  04} (2010) 015}, \href{http://arxiv.org/abs/0912.5463}{{\ttfamily
  arXiv:0912.5463 [hep-ph]}}.

\bibitem{Hertzberg:2010dc}
M.~P. Hertzberg, ``{On Inflation with Non-minimal Coupling},''
  \href{http://dx.doi.org/10.1007/JHEP11(2010)023}{{\em JHEP} {\bfseries 11}
  (2010) 023}, \href{http://arxiv.org/abs/1002.2995}{{\ttfamily arXiv:1002.2995
  [hep-ph]}}.

\bibitem{Burgess:2010zq}
C.~P. Burgess, H.~M. Lee, and M.~Trott, ``{Comment on Higgs Inflation and
  Naturalness},'' \href{http://dx.doi.org/10.1007/JHEP07(2010)007}{{\em JHEP}
  {\bfseries 07} (2010) 007}, \href{http://arxiv.org/abs/1002.2730}{{\ttfamily
  arXiv:1002.2730 [hep-ph]}}.

\bibitem{Chen:2009zp}
X.~Chen and Y.~Wang, ``{Quasi-Single Field Inflation and Non-Gaussianities},''
  \href{http://dx.doi.org/10.1088/1475-7516/2010/04/027}{{\em JCAP} {\bfseries
  04} (2010) 027}, \href{http://arxiv.org/abs/0911.3380}{{\ttfamily
  arXiv:0911.3380 [hep-th]}}.

\bibitem{Chung:2004nh}
D.~J.~H. Chung, E.~W. Kolb, A.~Riotto, and L.~Senatore, ``{Isocurvature
  constraints on gravitationally produced superheavy dark matter},''
  \href{http://dx.doi.org/10.1103/PhysRevD.72.023511}{{\em Phys. Rev. D}
  {\bfseries 72} (2005) 023511},
  \href{http://arxiv.org/abs/astro-ph/0411468}{{\ttfamily
  arXiv:astro-ph/0411468}}.

\bibitem{Garcia:2023qab}
M.~A.~G. Garcia, M.~Pierre, and S.~Verner, ``{New window into gravitationally
  produced scalar dark matter},''
  \href{http://dx.doi.org/10.1103/PhysRevD.108.115024}{{\em Phys. Rev. D}
  {\bfseries 108} no.~11, (2023) 115024},
  \href{http://arxiv.org/abs/2305.14446}{{\ttfamily arXiv:2305.14446
  [hep-ph]}}.

\bibitem{Liddle:1999pr}
A.~R. Liddle and A.~Mazumdar, ``{Perturbation amplitude in isocurvature
  inflation scenarios},''
  \href{http://dx.doi.org/10.1103/PhysRevD.61.123507}{{\em Phys. Rev. D}
  {\bfseries 61} (2000) 123507},
  \href{http://arxiv.org/abs/astro-ph/9912349}{{\ttfamily
  arXiv:astro-ph/9912349}}.

\bibitem{Li:2019ves}
L.~Li, T.~Nakama, C.~M. Sou, Y.~Wang, and S.~Zhou, ``{Gravitational Production
  of Superheavy Dark Matter and Associated Cosmological Signatures},''
  \href{http://dx.doi.org/10.1007/JHEP07(2019)067}{{\em JHEP} {\bfseries 07}
  (2019) 067}, \href{http://arxiv.org/abs/1903.08842}{{\ttfamily
  arXiv:1903.08842 [astro-ph.CO]}}.

\bibitem{Sou:2021juh}
C.~M. Sou, X.~Tong, and Y.~Wang, ``{Chemical-potential-assisted particle
  production in FRW spacetimes},''
  \href{http://dx.doi.org/10.1007/JHEP06(2021)129}{{\em JHEP} {\bfseries 06}
  (2021) 129}, \href{http://arxiv.org/abs/2104.08772}{{\ttfamily
  arXiv:2104.08772 [hep-th]}}.

\bibitem{Linde:2001ae}
A.~D. Linde, ``{Fast roll inflation},''
  \href{http://dx.doi.org/10.1088/1126-6708/2001/11/052}{{\em JHEP} {\bfseries
  11} (2001) 052}, \href{http://arxiv.org/abs/hep-th/0110195}{{\ttfamily
  arXiv:hep-th/0110195}}.

\bibitem{Co:2022aav}
R.~T. Co, T.~Gherghetta, and K.~Harigaya, ``{Axiogenesis with a heavy QCD
  axion},'' \href{http://dx.doi.org/10.1007/JHEP10(2022)121}{{\em JHEP}
  {\bfseries 10} (2022) 121}, \href{http://arxiv.org/abs/2206.00678}{{\ttfamily
  arXiv:2206.00678 [hep-ph]}}.

\bibitem{Linde:1996gt}
A.~D. Linde and V.~F. Mukhanov, ``{Nongaussian isocurvature perturbations from
  inflation},'' \href{http://dx.doi.org/10.1103/PhysRevD.56.R535}{{\em Phys.
  Rev. D} {\bfseries 56} (1997) R535--R539},
  \href{http://arxiv.org/abs/astro-ph/9610219}{{\ttfamily
  arXiv:astro-ph/9610219}}.

\bibitem{Enqvist:2001zp}
K.~Enqvist and M.~S. Sloth, ``{Adiabatic CMB perturbations in pre - big bang
  string cosmology},''
  \href{http://dx.doi.org/10.1016/S0550-3213(02)00043-3}{{\em Nucl. Phys. B}
  {\bfseries 626} (2002) 395--409},
  \href{http://arxiv.org/abs/hep-ph/0109214}{{\ttfamily arXiv:hep-ph/0109214}}.

\bibitem{Lyth:2001nq}
D.~H. Lyth and D.~Wands, ``{Generating the curvature perturbation without an
  inflaton},'' \href{http://dx.doi.org/10.1016/S0370-2693(01)01366-1}{{\em
  Phys. Lett. B} {\bfseries 524} (2002) 5--14},
  \href{http://arxiv.org/abs/hep-ph/0110002}{{\ttfamily arXiv:hep-ph/0110002}}.

\bibitem{Moroi:2001ct}
T.~Moroi and T.~Takahashi, ``{Effects of cosmological moduli fields on cosmic
  microwave background},''
  \href{http://dx.doi.org/10.1016/S0370-2693(01)01295-3}{{\em Phys. Lett. B}
  {\bfseries 522} (2001) 215--221},
  \href{http://arxiv.org/abs/hep-ph/0110096}{{\ttfamily arXiv:hep-ph/0110096}}.
  [Erratum: Phys.Lett.B 539, 303--303 (2002)].

\bibitem{Tkachev:1995md}
I.~I. Tkachev, ``{Phase transitions at preheating},''
  \href{http://dx.doi.org/10.1016/0370-2693(96)00297-3}{{\em Phys. Lett. B}
  {\bfseries 376} (1996) 35--40},
  \href{http://arxiv.org/abs/hep-th/9510146}{{\ttfamily arXiv:hep-th/9510146}}.

\bibitem{Kofman:1995fi}
L.~Kofman, A.~D. Linde, and A.~A. Starobinsky, ``{Nonthermal phase transitions
  after inflation},'' \href{http://dx.doi.org/10.1103/PhysRevLett.76.1011}{{\em
  Phys. Rev. Lett.} {\bfseries 76} (1996) 1011--1014},
  \href{http://arxiv.org/abs/hep-th/9510119}{{\ttfamily arXiv:hep-th/9510119}}.

\bibitem{Kasuya:1996ns}
S.~Kasuya, M.~Kawasaki, and T.~Yanagida, ``{Cosmological axion problem in
  chaotic inflationary universe},''
  \href{http://dx.doi.org/10.1016/S0370-2693(97)00809-5}{{\em Phys. Lett. B}
  {\bfseries 409} (1997) 94--100},
  \href{http://arxiv.org/abs/hep-ph/9608405}{{\ttfamily arXiv:hep-ph/9608405}}.

\bibitem{Kasuya:1997ha}
S.~Kasuya and M.~Kawasaki, ``{Can topological defects be formed during
  preheating?},'' \href{http://dx.doi.org/10.1103/PhysRevD.56.7597}{{\em Phys.
  Rev. D} {\bfseries 56} (1997) 7597--7607},
  \href{http://arxiv.org/abs/hep-ph/9703354}{{\ttfamily arXiv:hep-ph/9703354}}.

\bibitem{Kasuya:1998td}
S.~Kasuya and M.~Kawasaki, ``{Topological defects formation after inflation on
  lattice simulation},''
  \href{http://dx.doi.org/10.1103/PhysRevD.58.083516}{{\em Phys. Rev. D}
  {\bfseries 58} (1998) 083516},
  \href{http://arxiv.org/abs/hep-ph/9804429}{{\ttfamily arXiv:hep-ph/9804429}}.

\bibitem{Tkachev:1998dc}
I.~Tkachev, S.~Khlebnikov, L.~Kofman, and A.~D. Linde, ``{Cosmic strings from
  preheating},'' \href{http://dx.doi.org/10.1016/S0370-2693(98)01094-6}{{\em
  Phys. Lett. B} {\bfseries 440} (1998) 262--268},
  \href{http://arxiv.org/abs/hep-ph/9805209}{{\ttfamily arXiv:hep-ph/9805209}}.

\bibitem{Kasuya:1999hy}
S.~Kasuya and M.~Kawasaki, ``{Comments on cosmic string formation during
  preheating on lattice simulations},''
  \href{http://dx.doi.org/10.1103/PhysRevD.61.083510}{{\em Phys. Rev. D}
  {\bfseries 61} (2000) 083510},
  \href{http://arxiv.org/abs/hep-ph/9903324}{{\ttfamily arXiv:hep-ph/9903324}}.

\bibitem{Ballesteros:2021bee}
G.~Ballesteros, A.~Ringwald, C.~Tamarit, and Y.~Welling, ``{Revisiting
  isocurvature bounds in models unifying the axion with the inflaton},''
  \href{http://dx.doi.org/10.1088/1475-7516/2021/09/036}{{\em JCAP} {\bfseries
  09} (2021) 036}, \href{http://arxiv.org/abs/2104.13847}{{\ttfamily
  arXiv:2104.13847 [hep-ph]}}.

\bibitem{Georgi:1981pu}
H.~M. Georgi, L.~J. Hall, and M.~B. Wise, ``{Grand Unified Models With an
  Automatic {Peccei-Quinn} Symmetry},''
  \href{http://dx.doi.org/10.1016/0550-3213(81)90433-8}{{\em Nucl. Phys. B}
  {\bfseries 192} (1981) 409--416}.

\bibitem{Dine:1986bg}
M.~Dine and N.~Seiberg, ``{String Theory and the Strong {CP} Problem},''
  \href{http://dx.doi.org/10.1016/0550-3213(86)90043-X}{{\em Nucl. Phys. B}
  {\bfseries 273} (1986) 109--124}.

\bibitem{Coleman:1989zu}
S.~R. Coleman and K.-M. Lee, ``{WORMHOLES MADE WITHOUT MASSLESS MATTER
  FIELDS},'' \href{http://dx.doi.org/10.1016/0550-3213(90)90149-8}{{\em Nucl.
  Phys. B} {\bfseries 329} (1990) 387--409}.

\bibitem{Abbott:1989jw}
L.~F. Abbott and M.~B. Wise, ``{Wormholes and Global Symmetries},''
  \href{http://dx.doi.org/10.1016/0550-3213(89)90503-8}{{\em Nucl. Phys. B}
  {\bfseries 325} (1989) 687--704}.

\bibitem{Holman:1992us}
R.~Holman, S.~D.~H. Hsu, T.~W. Kephart, E.~W. Kolb, R.~Watkins, and L.~M.
  Widrow, ``{Solutions to the strong CP problem in a world with gravity},''
  \href{http://dx.doi.org/10.1016/0370-2693(92)90491-L}{{\em Phys. Lett. B}
  {\bfseries 282} (1992) 132--136},
  \href{http://arxiv.org/abs/hep-ph/9203206}{{\ttfamily arXiv:hep-ph/9203206}}.

\bibitem{Kamionkowski:1992mf}
M.~Kamionkowski and J.~March-Russell, ``{Planck scale physics and the
  Peccei-Quinn mechanism},''
  \href{http://dx.doi.org/10.1016/0370-2693(92)90492-M}{{\em Phys. Lett. B}
  {\bfseries 282} (1992) 137--141},
  \href{http://arxiv.org/abs/hep-th/9202003}{{\ttfamily arXiv:hep-th/9202003}}.

\bibitem{Barr:1992qq}
S.~M. Barr and D.~Seckel, ``{Planck scale corrections to axion models},''
  \href{http://dx.doi.org/10.1103/PhysRevD.46.539}{{\em Phys. Rev. D}
  {\bfseries 46} (1992) 539--549}.

\bibitem{Ghigna:1992iv}
S.~Ghigna, M.~Lusignoli, and M.~Roncadelli, ``{Instability of the invisible
  axion},'' \href{http://dx.doi.org/10.1016/0370-2693(92)90019-Z}{{\em Phys.
  Lett. B} {\bfseries 283} (1992) 278--281}.

\bibitem{Alonso:2017avz}
R.~Alonso and A.~Urbano, ``{Wormholes and masses for Goldstone bosons},''
  \href{http://dx.doi.org/10.1007/JHEP02(2019)136}{{\em JHEP} {\bfseries 02}
  (2019) 136}, \href{http://arxiv.org/abs/1706.07415}{{\ttfamily
  arXiv:1706.07415 [hep-ph]}}.

\bibitem{Kallosh:1995hi}
R.~Kallosh, A.~D. Linde, D.~A. Linde, and L.~Susskind, ``{Gravity and global
  symmetries},'' \href{http://dx.doi.org/10.1103/PhysRevD.52.912}{{\em Phys.
  Rev. D} {\bfseries 52} (1995) 912--935},
  \href{http://arxiv.org/abs/hep-th/9502069}{{\ttfamily arXiv:hep-th/9502069}}.

\bibitem{Vafa:1984xg}
C.~Vafa and E.~Witten, ``{Parity Conservation in QCD},''
  \href{http://dx.doi.org/10.1103/PhysRevLett.53.535}{{\em Phys. Rev. Lett.}
  {\bfseries 53} (1984) 535}.

\bibitem{Georgi:1986kr}
H.~Georgi and L.~Randall, ``{Flavor Conserving CP Violation in Invisible Axion
  Models},'' \href{http://dx.doi.org/10.1016/0550-3213(86)90022-2}{{\em Nucl.
  Phys. B} {\bfseries 276} (1986) 241--252}.

\bibitem{Gorghetto:2018ocs}
M.~Gorghetto and G.~Villadoro, ``{Topological Susceptibility and QCD Axion
  Mass: QED and NNLO corrections},''
  \href{http://dx.doi.org/10.1007/JHEP03(2019)033}{{\em JHEP} {\bfseries 03}
  (2019) 033}, \href{http://arxiv.org/abs/1812.01008}{{\ttfamily
  arXiv:1812.01008 [hep-ph]}}.

\bibitem{Nakagawa:2020zjr}
S.~Nakagawa, F.~Takahashi, and M.~Yamada, ``{Trapping Effect for QCD Axion Dark
  Matter},'' \href{http://dx.doi.org/10.1088/1475-7516/2021/05/062}{{\em JCAP}
  {\bfseries 05} (2021) 062}, \href{http://arxiv.org/abs/2012.13592}{{\ttfamily
  arXiv:2012.13592 [hep-ph]}}.

\bibitem{DiLuzio:2021gos}
L.~Di~Luzio, B.~Gavela, P.~Quilez, and A.~Ringwald, ``{Dark matter from an even
  lighter QCD axion: trapped misalignment},''
  \href{http://dx.doi.org/10.1088/1475-7516/2021/10/001}{{\em JCAP} {\bfseries
  10} (2021) 001}, \href{http://arxiv.org/abs/2102.01082}{{\ttfamily
  arXiv:2102.01082 [hep-ph]}}.

\bibitem{Jeong:2022kdr}
K.~S. Jeong, K.~Matsukawa, S.~Nakagawa, and F.~Takahashi, ``{Cosmological
  effects of Peccei-Quinn symmetry breaking on QCD axion dark matter},''
  \href{http://dx.doi.org/10.1088/1475-7516/2022/03/026}{{\em JCAP} {\bfseries
  03} no.~03, (2022) 026}, \href{http://arxiv.org/abs/2201.00681}{{\ttfamily
  arXiv:2201.00681 [hep-ph]}}.

\bibitem{Buen-Abad:2019uoc}
M.~A. Buen-Abad and J.~Fan, ``{Dynamical axion misalignment with small
  instantons},'' \href{http://dx.doi.org/10.1007/JHEP12(2019)161}{{\em JHEP}
  {\bfseries 12} (2019) 161}, \href{http://arxiv.org/abs/1911.05737}{{\ttfamily
  arXiv:1911.05737 [hep-ph]}}.

\bibitem{Kibble_1976}
T.~W.~B. Kibble, ``Topology of cosmic domains and strings,''
  \href{http://dx.doi.org/10.1088/0305-4470/9/8/029}{{\em Journal of Physics A:
  Mathematical and General} {\bfseries 9} no.~8, (Aug, 1976) 1387--1398}.
  \url{https://doi.org/10.1088/0305-4470/9/8/029}.

\bibitem{PhysRevD.26.435}
T.~W.~B. Kibble, G.~Lazarides, and Q.~Shafi, ``Walls bounded by strings,''
  \href{http://dx.doi.org/10.1103/PhysRevD.26.435}{{\em Phys. Rev. D}
  {\bfseries 26} (Jul, 1982) 435--439}.
  \url{https://link.aps.org/doi/10.1103/PhysRevD.26.435}.

\bibitem{KIBBLE1980183}
T.~Kibble, ``Some implications of a cosmological phase transition,''
  \href{http://dx.doi.org/https://doi.org/10.1016/0370-1573(80)90091-5}{{\em
  Physics Reports} {\bfseries 67} no.~1, (1980) 183--199}.
  \url{https://www.sciencedirect.com/science/article/pii/0370157380900915}.

\bibitem{Linde:1990yj}
A.~D. Linde and D.~H. Lyth, ``{Axionic domain wall production during
  inflation},'' \href{http://dx.doi.org/10.1016/0370-2693(90)90613-B}{{\em
  Phys. Lett. B} {\bfseries 246} (1990) 353--358}.

\bibitem{Nelson:2018via}
A.~E. Nelson and H.~Xiao, ``{Axion Cosmology with Early Matter Domination},''
  \href{http://dx.doi.org/10.1103/PhysRevD.98.063516}{{\em Phys. Rev. D}
  {\bfseries 98} no.~6, (2018) 063516},
  \href{http://arxiv.org/abs/1807.07176}{{\ttfamily arXiv:1807.07176
  [astro-ph.CO]}}.

\bibitem{Co:2019jts}
R.~T. Co, L.~J. Hall, and K.~Harigaya, ``{Axion Kinetic Misalignment
  Mechanism},'' \href{http://dx.doi.org/10.1103/PhysRevLett.124.251802}{{\em
  Phys. Rev. Lett.} {\bfseries 124} no.~25, (2020) 251802},
  \href{http://arxiv.org/abs/1910.14152}{{\ttfamily arXiv:1910.14152
  [hep-ph]}}.

\bibitem{Spokoiny:1993kt}
B.~Spokoiny, ``{Deflationary universe scenario},''
  \href{http://dx.doi.org/10.1016/0370-2693(93)90155-B}{{\em Phys. Lett. B}
  {\bfseries 315} (1993) 40--45},
  \href{http://arxiv.org/abs/gr-qc/9306008}{{\ttfamily arXiv:gr-qc/9306008}}.

\bibitem{Joyce:1996cp}
M.~Joyce, ``{Electroweak Baryogenesis and the Expansion Rate of the
  Universe},'' \href{http://dx.doi.org/10.1103/PhysRevD.55.1875}{{\em Phys.
  Rev. D} {\bfseries 55} (1997) 1875--1878},
  \href{http://arxiv.org/abs/hep-ph/9606223}{{\ttfamily arXiv:hep-ph/9606223}}.

\bibitem{Ferreira:1997hj}
P.~G. Ferreira and M.~Joyce, ``{Cosmology with a primordial scaling field},''
  \href{http://dx.doi.org/10.1103/PhysRevD.58.023503}{{\em Phys. Rev. D}
  {\bfseries 58} (1998) 023503},
  \href{http://arxiv.org/abs/astro-ph/9711102}{{\ttfamily
  arXiv:astro-ph/9711102}}.

\bibitem{Co:2021lkc}
R.~T. Co, D.~Dunsky, N.~Fernandez, A.~Ghalsasi, L.~J. Hall, K.~Harigaya, and
  J.~Shelton, ``{Gravitational wave and CMB probes of axion kination},''
  \href{http://dx.doi.org/10.1007/JHEP09(2022)116}{{\em JHEP} {\bfseries 09}
  (2022) 116}, \href{http://arxiv.org/abs/2108.09299}{{\ttfamily
  arXiv:2108.09299 [hep-ph]}}.

\bibitem{Gouttenoire:2021wzu}
Y.~Gouttenoire, G.~Servant, and P.~Simakachorn, ``{Revealing the Primordial
  Irreducible Inflationary Gravitational-Wave Background with a Spinning
  Peccei-Quinn Axion},'' \href{http://arxiv.org/abs/2108.10328}{{\ttfamily
  arXiv:2108.10328 [hep-ph]}}.

\bibitem{Gouttenoire:2021jhk}
Y.~Gouttenoire, G.~Servant, and P.~Simakachorn, ``{Kination cosmology from
  scalar fields and gravitational-wave signatures},''
  \href{http://arxiv.org/abs/2111.01150}{{\ttfamily arXiv:2111.01150
  [hep-ph]}}.

\bibitem{Bettoni:2018utf}
D.~Bettoni and J.~Rubio, ``{Quintessential Affleck-Dine baryogenesis with
  non-minimal couplings},''
  \href{http://dx.doi.org/10.1016/j.physletb.2018.07.046}{{\em Phys. Lett. B}
  {\bfseries 784} (2018) 122--129},
  \href{http://arxiv.org/abs/1805.02669}{{\ttfamily arXiv:1805.02669
  [astro-ph.CO]}}.

\bibitem{Borsanyi:2016ksw}
S.~Borsanyi {\em et~al.}, ``{Calculation of the axion mass based on
  high-temperature lattice quantum chromodynamics},''
  \href{http://dx.doi.org/10.1038/nature20115}{{\em Nature} {\bfseries 539}
  no.~7627, (2016) 69--71}, \href{http://arxiv.org/abs/1606.07494}{{\ttfamily
  arXiv:1606.07494 [hep-lat]}}.

\bibitem{Hertzberg:2008wr}
M.~P. Hertzberg, M.~Tegmark, and F.~Wilczek, ``{Axion Cosmology and the Energy
  Scale of Inflation},''
  \href{http://dx.doi.org/10.1103/PhysRevD.78.083507}{{\em Phys. Rev. D}
  {\bfseries 78} (2008) 083507},
  \href{http://arxiv.org/abs/0807.1726}{{\ttfamily arXiv:0807.1726
  [astro-ph]}}.

\bibitem{Arias:2021rer}
P.~Arias, N.~Bernal, D.~Karamitros, C.~Maldonado, L.~Roszkowski, and
  M.~Venegas, ``{New opportunities for axion dark matter searches in
  nonstandard cosmological models},''
  \href{http://dx.doi.org/10.1088/1475-7516/2021/11/003}{{\em JCAP} {\bfseries
  11} (2021) 003}, \href{http://arxiv.org/abs/2107.13588}{{\ttfamily
  arXiv:2107.13588 [hep-ph]}}.

\bibitem{Arias:2022qjt}
P.~Arias, N.~Bernal, J.~K. Osi\'nski, and L.~Roszkowski, ``{Dark matter axions
  in the early universe with a period of increasing temperature},''
  \href{http://dx.doi.org/10.1088/1475-7516/2023/05/028}{{\em JCAP} {\bfseries
  05} (2023) 028}, \href{http://arxiv.org/abs/2207.07677}{{\ttfamily
  arXiv:2207.07677 [hep-ph]}}.

\bibitem{PhysRevLett.43.103}
J.~E. Kim, ``Weak-interaction singlet and strong $\mathrm{CP}$ invariance,''
  \href{http://dx.doi.org/10.1103/PhysRevLett.43.103}{{\em Phys. Rev. Lett.}
  {\bfseries 43} (Jul, 1979) 103--107}.
  \url{https://link.aps.org/doi/10.1103/PhysRevLett.43.103}.

\bibitem{SHIFMAN1980493}
M.~Shifman, A.~Vainshtein, and V.~Zakharov, ``Can confinement ensure natural cp
  invariance of strong interactions?,''
  \href{http://dx.doi.org/https://doi.org/10.1016/0550-3213(80)90209-6}{{\em
  Nuclear Physics B} {\bfseries 166} no.~3, (1980) 493--506}.
  \url{https://www.sciencedirect.com/science/article/pii/0550321380902096}.

\bibitem{Zhitnitsky:1980tq}
A.~R. Zhitnitsky, ``{On Possible Suppression of the Axion Hadron Interactions.
  (In Russian)},'' {\em Sov. J. Nucl. Phys.} {\bfseries 31} (1980) 260.

\bibitem{DINE1981199}
M.~Dine, W.~Fischler, and M.~Srednicki, ``A simple solution to the strong cp
  problem with a harmless axion,''
  \href{http://dx.doi.org/https://doi.org/10.1016/0370-2693(81)90590-6}{{\em
  Physics Letters B} {\bfseries 104} no.~3, (1981) 199--202}.
  \url{https://www.sciencedirect.com/science/article/pii/0370269381905906}.

\bibitem{RVolkas:2023jiv}
R.~R.~Volkas, ``{VISH\ensuremath{\nu}: Flavour-Variant DFSZ Axion Model for
  Inflation, Neutrino Masses, Dark Matter, and Baryogenesis},''
  \href{http://dx.doi.org/10.31526/lhep.2023.358}{{\em LHEP} {\bfseries 2023}
  (2023) 358}.

\bibitem{Kajantie:2002wa}
K.~Kajantie, M.~Laine, K.~Rummukainen, and Y.~Schroder, ``{The Pressure of hot
  QCD up to g6 ln(1/g)},''
  \href{http://dx.doi.org/10.1103/PhysRevD.67.105008}{{\em Phys. Rev. D}
  {\bfseries 67} (2003) 105008},
  \href{http://arxiv.org/abs/hep-ph/0211321}{{\ttfamily arXiv:hep-ph/0211321}}.

\bibitem{Davoudiasl:2004gf}
H.~Davoudiasl, R.~Kitano, G.~D. Kribs, H.~Murayama, and P.~J. Steinhardt,
  ``{Gravitational baryogenesis},''
  \href{http://dx.doi.org/10.1103/PhysRevLett.93.201301}{{\em Phys. Rev. Lett.}
  {\bfseries 93} (2004) 201301},
  \href{http://arxiv.org/abs/hep-ph/0403019}{{\ttfamily arXiv:hep-ph/0403019}}.

\bibitem{Linde:1996cx}
A.~D. Linde, ``{Relaxing the cosmological moduli problem},''
  \href{http://dx.doi.org/10.1103/PhysRevD.53.R4129}{{\em Phys. Rev. D}
  {\bfseries 53} (1996) R4129--R4132},
  \href{http://arxiv.org/abs/hep-th/9601083}{{\ttfamily arXiv:hep-th/9601083}}.

\bibitem{Deur:2022msf}
A.~Deur, V.~Burkert, J.~P. Chen, and W.~Korsch, ``{Experimental determination
  of the QCD effective charge $\alpha_{g_1}(Q)$},''
  \href{http://dx.doi.org/10.3390/particles5020015}{{\em Particles} {\bfseries
  5} (2022) 171}, \href{http://arxiv.org/abs/2205.01169}{{\ttfamily
  arXiv:2205.01169 [hep-ph]}}.

\bibitem{Caldwell:2013mox}
R.~R. Caldwell and S.~S. Gubser, ``{Brief history of curvature},''
  \href{http://dx.doi.org/10.1103/PhysRevD.87.063523}{{\em Phys. Rev. D}
  {\bfseries 87} no.~6, (2013) 063523},
  \href{http://arxiv.org/abs/1302.1201}{{\ttfamily arXiv:1302.1201
  [astro-ph.CO]}}.

\bibitem{Huston:2023ofk}
J.~Huston, K.~Rabbertz, and G.~Zanderighi, ``{Quantum Chromodynamics},''
  \href{http://arxiv.org/abs/2312.14015}{{\ttfamily arXiv:2312.14015
  [hep-ph]}}.

\bibitem{Wantz:2009it}
O.~Wantz and E.~P.~S. Shellard, ``{Axion Cosmology Revisited},''
  \href{http://dx.doi.org/10.1103/PhysRevD.82.123508}{{\em Phys. Rev. D}
  {\bfseries 82} (2010) 123508},
  \href{http://arxiv.org/abs/0910.1066}{{\ttfamily arXiv:0910.1066
  [astro-ph.CO]}}.

\bibitem{Linde:1987bx}
A.~D. Linde, ``{Inflation and Axion Cosmology},''
  \href{http://dx.doi.org/10.1016/0370-2693(88)90597-7}{{\em Phys. Lett. B}
  {\bfseries 201} (1988) 437--439}.

\bibitem{Wilczek:2004cr}
F.~Wilczek, ``{A Model of anthropic reasoning, addressing the dark to ordinary
  matter coincidence},'' \href{http://arxiv.org/abs/hep-ph/0408167}{{\ttfamily
  arXiv:hep-ph/0408167}}.

\bibitem{Tegmark:2005dy}
M.~Tegmark, A.~Aguirre, M.~Rees, and F.~Wilczek, ``{Dimensionless constants,
  cosmology and other dark matters},''
  \href{http://dx.doi.org/10.1103/PhysRevD.73.023505}{{\em Phys. Rev. D}
  {\bfseries 73} (2006) 023505},
  \href{http://arxiv.org/abs/astro-ph/0511774}{{\ttfamily
  arXiv:astro-ph/0511774}}.

\bibitem{Alonso-Alvarez:2018tus}
G.~Alonso-\'Alvarez and J.~Jaeckel, ``{Lightish but clumpy: scalar dark matter
  from inflationary fluctuations},''
  \href{http://dx.doi.org/10.1088/1475-7516/2018/10/022}{{\em JCAP} {\bfseries
  10} (2018) 022}, \href{http://arxiv.org/abs/1807.09785}{{\ttfamily
  arXiv:1807.09785 [hep-ph]}}.

\bibitem{Herranen:2015ima}
M.~Herranen, T.~Markkanen, S.~Nurmi, and A.~Rajantie, ``{Spacetime curvature
  and Higgs stability after inflation},''
  \href{http://dx.doi.org/10.1103/PhysRevLett.115.241301}{{\em Phys. Rev.
  Lett.} {\bfseries 115} (2015) 241301},
  \href{http://arxiv.org/abs/1506.04065}{{\ttfamily arXiv:1506.04065
  [hep-ph]}}.

\bibitem{Bassett:1997az}
B.~A. Bassett and S.~Liberati, ``{Geometric reheating after inflation},''
  \href{http://dx.doi.org/10.1103/PhysRevD.60.049902}{{\em Phys. Rev. D}
  {\bfseries 58} (1998) 021302},
  \href{http://arxiv.org/abs/hep-ph/9709417}{{\ttfamily arXiv:hep-ph/9709417}}.
  [Erratum: Phys.Rev.D 60, 049902 (1999)].

\bibitem{Dufaux:2006ee}
J.~F. Dufaux, G.~N. Felder, L.~Kofman, M.~Peloso, and D.~Podolsky,
  ``{Preheating with trilinear interactions: Tachyonic resonance},''
  \href{http://dx.doi.org/10.1088/1475-7516/2006/07/006}{{\em JCAP} {\bfseries
  07} (2006) 006}, \href{http://arxiv.org/abs/hep-ph/0602144}{{\ttfamily
  arXiv:hep-ph/0602144}}.

\bibitem{Markkanen:2015xuw}
T.~Markkanen and S.~Nurmi, ``{Dark matter from gravitational particle
  production at reheating},''
  \href{http://dx.doi.org/10.1088/1475-7516/2017/02/008}{{\em JCAP} {\bfseries
  02} (2017) 008}, \href{http://arxiv.org/abs/1512.07288}{{\ttfamily
  arXiv:1512.07288 [astro-ph.CO]}}.

\bibitem{Chung:2018ayg}
D.~J.~H. Chung, E.~W. Kolb, and A.~J. Long, ``{Gravitational production of
  super-Hubble-mass particles: an analytic approach},''
  \href{http://dx.doi.org/10.1007/JHEP01(2019)189}{{\em JHEP} {\bfseries 01}
  (2019) 189}, \href{http://arxiv.org/abs/1812.00211}{{\ttfamily
  arXiv:1812.00211 [hep-ph]}}.

\bibitem{Bartrum:2013fia}
S.~Bartrum, M.~Bastero-Gil, A.~Berera, R.~Cerezo, R.~O. Ramos, and J.~G. Rosa,
  ``{The importance of being warm (during inflation)},''
  \href{http://dx.doi.org/10.1016/j.physletb.2014.03.029}{{\em Phys. Lett. B}
  {\bfseries 732} (2014) 116--121},
  \href{http://arxiv.org/abs/1307.5868}{{\ttfamily arXiv:1307.5868 [hep-ph]}}.

\bibitem{Babichev:2020xeg}
E.~Babichev, D.~Gorbunov, and S.~Ramazanov, ``{Gravitational misalignment
  mechanism of Dark Matter production},''
  \href{http://dx.doi.org/10.1088/1475-7516/2020/08/047}{{\em JCAP} {\bfseries
  08} (2020) 047}, \href{http://arxiv.org/abs/2004.03410}{{\ttfamily
  arXiv:2004.03410 [hep-ph]}}.

\end{thebibliography}\endgroup

\end{document}